\documentclass[a4paper,twocolumn,10pt, accepted = 2026-07-27]{quantumarticle}

\pdfoutput=1
\usepackage[utf8]{inputenc}
\usepackage[english]{babel}
\usepackage[T1]{fontenc}
\usepackage{amsmath}
\usepackage{hyperref}

\usepackage{lipsum}

\usepackage{graphicx}
\usepackage{dcolumn}
\usepackage{bm}

\usepackage{multirow}

\usepackage{amsmath,amssymb,amsfonts}
\usepackage{graphicx}
\usepackage{color}
\usepackage{appendix}
\usepackage{amsthm}
\usepackage{tikz}
\usepackage{youngtab}
\usepackage{mleftright}
\usepackage{url}
\usepackage[most]{tcolorbox}
\allowdisplaybreaks

\usepackage{tikz}
\usetikzlibrary{shapes,snakes,positioning,automata,arrows.meta,calc,decorations.markings,math,arrows.meta}
\tikzstyle{vertex}=[circle, draw, inner sep=0pt, minimum size=6pt]
\usepackage{xcolor}
\usepackage{subfig}

\usepackage{mathtools}

\usepackage[export]{adjustbox}

\usepackage{geometry}
\usepackage{soul}

\usepackage[numbers,sort&compress]{natbib}

\newtheorem{definition}{Definition}

\newcommand{\nn}{\nonumber}

\def\a{\alpha}

\def\t{\tau}

\def\p{\psi}

\def\P{\Psi}

\def\<{\langle}
\def\>{\rangle}

\def\ha{{\hat{a}}}
\def\had{{\hat{a}^\dagger}}
\def\hb{{\hat{b}}}

\def\t2{{\tilde{1}}}

\usepackage[normalem]{ulem}

\newcommand\encircle[1]{%
	\tikz[baseline=(X.base)] 
	\node (X) [draw, shape=circle, inner sep=0] {\strut #1};}

\makeatletter
\providecommand\UseOneTimeHook[1]{}%
\makeatother

\begin{document}

\title{Efficient Graph State Generation in Linear Optics}

	\author{Seungbeom Chin}
	\email{sbthesy@gmail.com}
	\affiliation{Okinawa Institute of Science and Technology Graduate University, Okinawa 904-0495, Japan}

    	\author{William J. Munro}
        \email{bill.munro@oist.jp}
	\affiliation{Okinawa Institute of Science and Technology Graduate University, Okinawa 904-0495, Japan}

\maketitle

\begin{abstract}
 Graph states are central resources for quantum information processing, supporting applications in computation, communication, and error correction. In photonic systems, they are typically assembled from smaller entangled states using probabilistic fusion gates, which demand many photons and suffer from low success rates. We present an optimized scheme for directly generating caterpillar graph states (CGSs)---essential resource states for constructing high-dimensional lattice graph states---using only single-photon sources, linear optics, and heralded measurements. Based on the linear quantum graph (LQG) picture, our method produces CGSs efficiently. For CGSs of length $l\ge 3$, it requires $l-2$ fewer photons and achieves a success rate $2^{l-2}$ times higher than fusion-based approaches. These results demonstrate that the LQG picture provides a powerful and flexible route to generating complex photonic graph states for efficient quantum information processing.
\end{abstract}

\section{Introduction}
    
Quantum technologies promise to transform how we process, transmit, and secure information. Among the most promising platforms is photonic quantum information processing~\cite{knill2001scheme,kok2007linear,obrien2007,azuma2015all,azuma2015allqkd}, which exploits the quantum properties of light for computation and communication. Photons are fast, robust against decoherence, and compatible with existing optical networks. However, scaling such systems remains challenging---particularly in the efficient creation of large entangled resource states. A powerful class of such resources is graph states~\cite{raussendorf2001one,raussendorf2003measurement,hein2004multiparty}, where vertices represent qubits and edges denote entangling operations. Graph states underpin measurement-based quantum computation (MBQC)~\cite{briegel2009measurement,van2013universal}, quantum error correction~\cite{hein2006entanglement,schlingemann}, quantum secret sharing~\cite{markham2008,bell2014experimental}, and studies of quantum nonlocality~\cite{coiteux2025genuinely}. In photonic systems, they are typically constructed using fusion gates~\cite{browne2005resource,varnava2006loss,varnava2008good,li2015resource,bartolucci2023fusion,lee2023graph}, which probabilistically combine smaller entangled states. Although conceptually simple, fusion-based methods require many entangled photon pairs and suffer from exponentially decreasing success rates.

An alternative strategy is to generate entanglement directly from single-photon sources using only linear optics and heralded measurements. A promising framework for such an approach is the \emph{linear quantum graph (LQG) picture}~\cite{chin2021graph,chin2024shortcut,chin2024heralded,chin2024exponentially}, which allows systematic design of entanglement schemes through graph-theoretic methods. In this work, we extend the LQG framework to introduce an optimized method for generating caterpillar graph states (CGSs). They serve as fundamental building blocks for the  scalable construction of high-dimensional lattice graph states by Type-I fusions~\cite{hilaire2023near,pettersson2025deterministic} and are used for the certification of robust genuinely multipartite nonlocality~\cite{coiteux2025genuinely}. Using two key structures---path circuit graphs and primate circuit  graphs---we construct optical circuits that generate CGSs with fewer photons and significantly higher success probabilities than fusion-based approaches. Our approach demonstrates the advantage of employing a broader set of measurement gates beyond conventional fusion gates.
It enables efficient generation of arbitrarily high-dimensional cluster states in MBQC and provides an optimal platform 
for testing local-operations-and-shared-randomness genuine multipartite nonlocality (LOSR-GMNL)~\cite{coiteux2025genuinely}.

The remainder of this paper is organized as follows. Section II reviews the LQG picture and two equivalent representations in it---the undirected bipartite and directed unipartite representations. Section III introduces the subgraph structures underlying CGS generation. Section IV presents explicit optical circuits based on the graph structure, and Section V concludes with scalability and outlook.

\section{LQG picture: Directed unipartite graph representation}\label{sec:lqg}

\begin{table*}[t]
\centering 
			\begin{tabular}{|l|l|l|l|}
				\hline
				\textbf{Sculpting operators} & \textbf{Sculpting bigraphs}  & \textbf{Sculpting digraphs}  \\
				\textbf{in bosonic systems} & ~~~~~in \textbf{$G_{ub} =(U\cup V, E)$}& ~~~~~in \textbf{$G_{du} =(W, E)$} \\
				\hline\hline 
				Spatial modes $j$ & Labelled circles (\encircle{$j$}) $\in$ $U$ & Circles (\encircle{$j$}) $\in$ $W$    \\ \hline 
				$\hat{A}^{(l)}$  
    &     Unlabelled dots ($\bullet$) $\in$  $V$  & Circles (\encircle{$l$}) $\in$  $W$  \\ \hline 
				Spatial distributions of $\hat{A}^{(l)}$  & Undirected edges  $\in$ $E$ & Directed edges  $\in$ $E$ \\ \hline 
				Probability amplitude $\a_j^{(l)} $ & Edge weight $\a_j^{(l)} $ & Edge weight $\a_j^{(l)} $\\ \hline 
				Internal state $\p_j^{(l)} $ & Edge weight $\p_j^{(l)}$  & Edge weight $\p_j^{(l)}$\\
				\hline 
			\end{tabular}
			\caption{Correspondence relations between sculpting operators and sculpting bigraphs. Note that in $G_{du}$, both spatical modes and subtraction operators are mapped to circles in $W$. 
            Because we only work on internal states $\{|+\>,|-\>, |0\>,|1\>\}$  $\big(|0\> = \frac{1}{\sqrt{2}}(|+\> + |-\>)$, $|1\> = \frac{1}{\sqrt{2}}(|+\> -|-\>) \big)$, the edge weight for the internal states are expressed as edge colors \{Solid Black, Dashed Black, Red, Blue\} for simplicity. Hence, when we mention edge weights (colors) from now on, they are  the edge  weights for probability amplitudes (internal states).}
            			\label{dict}
		\end{table*}

The LQG picture provides a systematic framework for designing heralded photonic entanglement schemes. 
Intuitively, a sculpting protocol generates entanglement by coherently subtracting single photons across multiple spatial modes. The LQG picture translates these operations into graph structures, where vertices represent modes and edges encode how photon subtraction processes are superposed.
Our design follows two conceptual steps:

First, the heralding process is described by the sculpting protocol, which generates entangled states through spatially overlapped single-boson annihilation operators, referred to as sculpting operators. Sculpting operators are single-boson annihilation operators that are superposed across different spatial modes. 
See Appendix B for a detailed explanation of the sculpting protocol. A major technical difficulty in generating specific target states lies in identifying appropriate sculpting operators. The LQG picture was proposed to solve this problem systematically, with each sculpting operator mapped to a sculpting bigraph~\cite{chin2024shortcut}. 

Second, to achieve experimentally realizable linear optical circuits, we need to convert  the bigraph elements into linear-optical elements. We use the
translation rules introduced in Ref.~\cite{chin2024heralded}, which provide a straightforward conversion from graphs to circuits. Therefore, \emph{once an appropriate sculpting bigraph is identified, a corresponding heralded optical network can be constructed directly.    
}

In this section, we explain two equivalent graph representations in the LQG picture that describe sculpting protocols: the undirected bipartite graph (bigraph) representation (denoted as $G_{ub}$) and the directed unipartite graph representation (denoted as $G_{du}$)~\cite{chin2021graph}.  
While $G_{ub}$ is advantageous for intuitively describing sculpting protocols, $G_{du}$ is more suitable for displaying complex entanglement structures, including our graph solutions. Each representation is employed wherever it is more convenient for the discussion.


\paragraph*{Undirected bipartite representation $G_{ub}$.---}
	The graph mapping of a sculpting operator $\hat{A}_{N+K}$ (see Eq.~\eqref{annihilation} of Appendix~\ref{sculpting} for the definition) in $G_{ub}$ is enumerated in Table I. 
	In the bigraph picture, the sculpting bigraphs are arranged such that the final state terms in \eqref{final_state} of Appendix~\ref{sculpting} correspond to the  perfect matchings (PMs) of the sculpting bigraph  (the no-bunching condition~\cite{chin2024shortcut}). And it is generally difficult to determine suitable edge weights for a sculpting bigraph to both satisfy this property and generate genuine multipartite entanglement (GME, entanglement inseparable across every bipartition). To resolve this issue, a convenient type of bigraphs that automatically fulfill the no-bunching condition was introduced~\cite{chin2024shortcut}. They are named \emph{effective perfect matching bigraphs} (EPM bigraphs), of which all the edges attach to the circles as one of the following forms
		\begin{align}
		&\begin{gathered}
				\includegraphics[width=.42\textwidth]{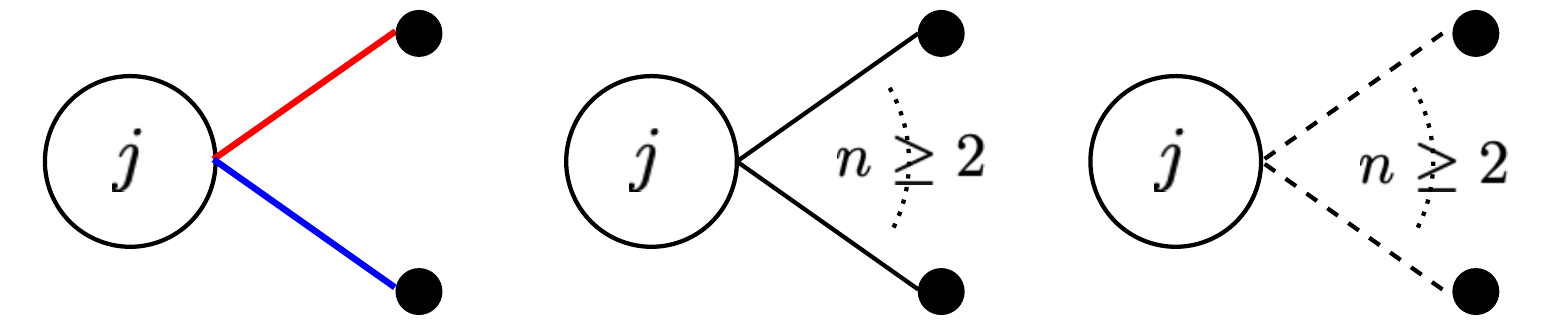}
			\end{gathered}, \nn 
		\end{align} 
        where the edge weights are all non-zero.

	If a sculpting bigraph is an EPM bigraph, then all the final state terms correspond to the PMs of the bigraph~\cite{chin2024shortcut}. EPM bigraphs are used to design schemes for generating various GME states. An example for the GHZ state is given in Fig.~\ref{fig:GHZ} of Appendix.

\paragraph*{Directed unipartite representation $G_{du}$.---} 
A digraph from a given bigraph is obtained by merging one circle in $U$ and one dot in $V$ and impose a direction from the dot to the circle to preserve the distinction between dots and circles (or more rigorously, from Definition~\ref{def:bi_to_di} of Appendix~\ref{glossary}). Then the digraph mapping of sculpting operators is given as Table~I.
For the remainder of this paper, a digraph that corresponds to a sculpting bigraph, hence to a sculpting operator, is called a \emph{sculpting digraph}. 
From the definition of EPM bigraphs and Table~I, we can see that \emph{EPM digraphs} are those whose edges come to the circles as one of the following forms
     		\begin{align}
			\begin{gathered}
		\includegraphics[width=.42\textwidth]{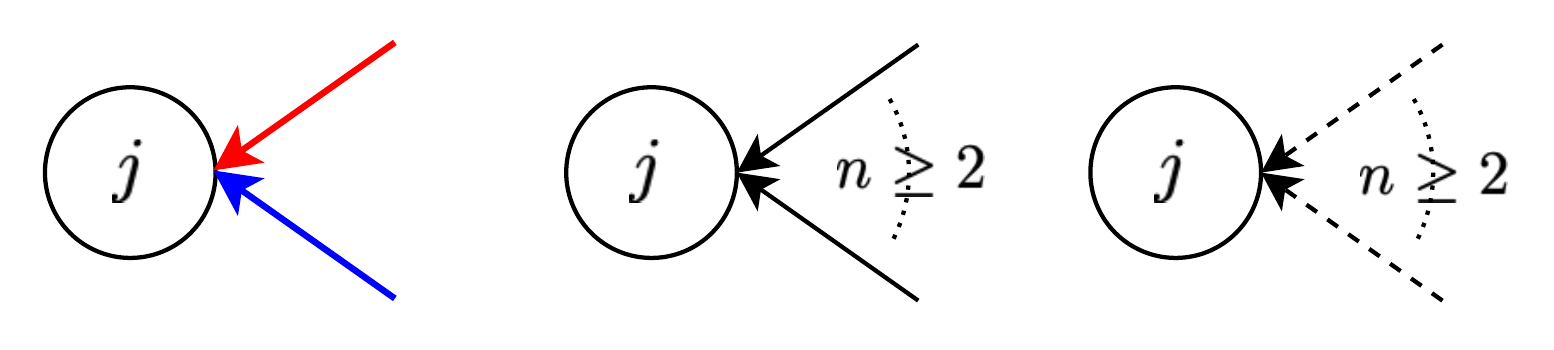}
			\end{gathered}. \nn 
		\end{align} 
	
There are two useful properties that sculpting digraphs have: First, in a sculpting digraph, a cycle (a closed path other than loops) corresponds to an independent PM in the bigraph counterpart. Therefore, when a bigraph has a complicated structure in a large system, the digraph representation is much easier to verify the PMs~\cite{chin2021graph}. 
And one PM in the bigraph representation is expressed as a \emph{disjoint cycle cover (DCC)} in $G_{ud}$ (see Appendix A for the definition). 

Second, we can find necessary conditions for a sculpting operator to generate genuine multipartite entanglement with the digraph representation. For a sculpting operator to generate a GME state, 
        all the vertices in it must be strongly connected to each other. This property can be an variation of Theorem 1 in Ref.~\cite{chin2021graph}. 
This necessary condition provides a powerful guideline to search for sculpting operators to generate GME states. We can find such operators by selecting strongly connected EPM digraphs as candidates and checking the final state by verifying their DCCs, which significantly limits the number of cases to be considered. 
 The digraph representation $G_{du}$ is particularly useful for the schemes in this work whose
 sculpting operators have complicated ancillary structures, as we will see in the next section.

	\section{Sculpting digraphs for caterpillar graph states}\label{sec:sculpting_digraph}
	
	Caterpillar graph states (CGSs) have multipartite qubits that are connected with Controlled-Z (CZ) gates\footnote{Graph states are GME states whose qubits and their CZ gate interactions are represented as dots and edges respectively, hence its graph picture is different from LQG picture for different physical elements.} in the shape of a caterpillar, i.e., a tree graph in which every vertex is on the path or one edge away from the path. The most general form of a caterpillar graph with length $l$ is given by
    \begin{align}\label{caterpillar_arb}
        \includegraphics[width=.45\textwidth]{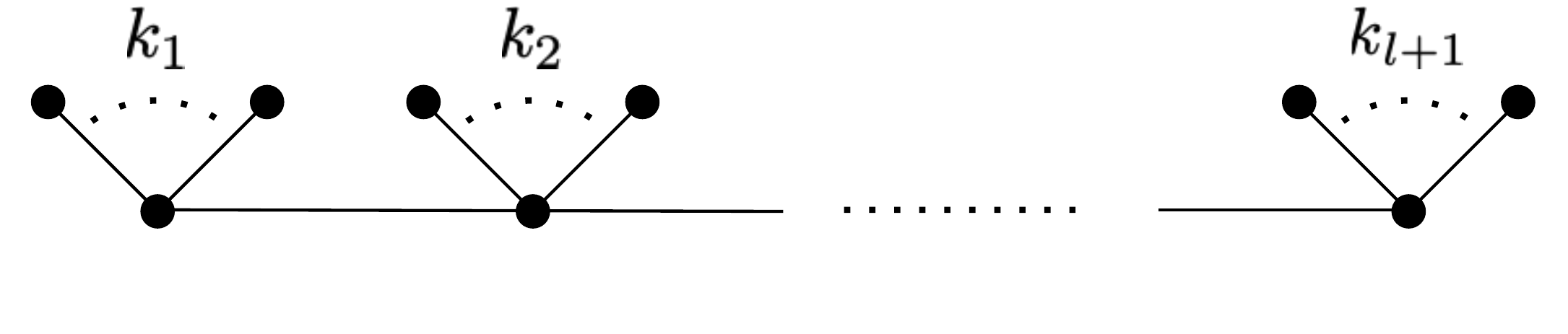}. 
    \end{align}
 A more detailed explanation on the CGSs is presented in Appendix~\ref{CGS}.
	
	Here we introduce a special type of sculpting digraphs that generate arbitrary CGSs. With our scheme, any CGS with a path of length $l$ is generated by the corresponding sculpting operator with $l+1$ ancillary particles in $l+1$ ancillary modes. The ancillary modes play the role of the path circuit that connects several star graphs, which we call the \emph{path circuit digraph} $P^{(l)}$.

\begin{figure*}
    \centering
         \includegraphics[width=1\textwidth]{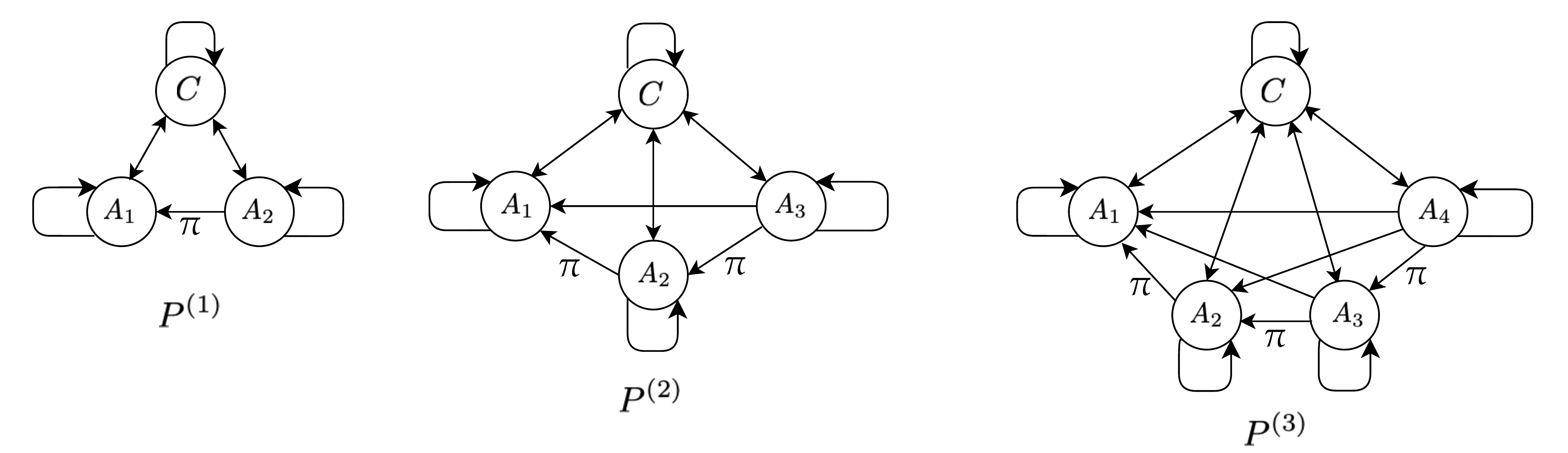}
        \caption{Path circuit digraphs $P^{(1)}$, $P^{(2)}$, and $P^{(3)}$. $P^{(l)}$ can be drawn from $P^{(l-1)}$ by adding $A_{l+1}$ and edges according to the iteration relation Eq.~\eqref{central_path_adj}.}
        \label{fig:central_path}
    \end{figure*}

A path circuit digraph $P^{(l)}$ of $(l+2)$ vertices $\{C,A_1,A_2,\cdots, A_n, A_{l+1}\}$ is a  digraph whose adjacency matrix is defined in an iterative manner as follows:
		\begin{align}\label{central_path_adj}
			P^{(1)} =& \mleft[
			\begin{array}{ccc}
				a_{CC} & a_{CA_1} & a_{CA_2} \\
				a_{A_1C} & a_{A_1A_1} & a_{A_1A_2} \\
				a_{A_2C} & a_{A_2A_1} & a_{A_2A_2}
			\end{array} \mright]
			= 
			\mleft[
			\begin{array}{ccc}
				1 & 1 & 1 \\
				1 & 1 & 0  \\
				1 & -1 & 1 \\
			\end{array}
			\mright]
			, \nn \\
			P^{(l)} =&
			\mleft[\begin{array}{ccccc}
				a_{CC} & a_{CA_1} & \cdots & a_{CA_{l+1}} \\
				a_{A_1C} & a_{A_1A_1} & \cdots & a_{A_1A_{l+1}} \\
				\vdots & \vdots  & \ddots &\vdots \\ 
				a_{A_{l+1}C} & a_{A_{l+1}A_1} & \cdots & a_{A_{l+1}A_{l+1}} 
			\end{array} \mright] \nn \\
			=&
			\mleft[
			\begin{array}{ccccc|c}
				&  &  & &  & 1 \\
				&  &  & &  & 0 \\
				&  &  & &  & \vdots \\  
				\multicolumn{5}{c|}{\smash{\raisebox{2\normalbaselineskip}{$P^{(l-1)}$}}} & 0 \\
				\hline 
				1 & 1 & \cdots & 1  & -1 & 1  \\
			\end{array}
			\mright]. 
		\end{align}	
	Digraphs corresponding to $P^{(1)}$, $P^{(2)}$, $P^{(3)}$ are shown in 
Fig.~\ref{fig:central_path}.

    \begin{figure*}[t]
    \centering
         \includegraphics[width=1\textwidth]{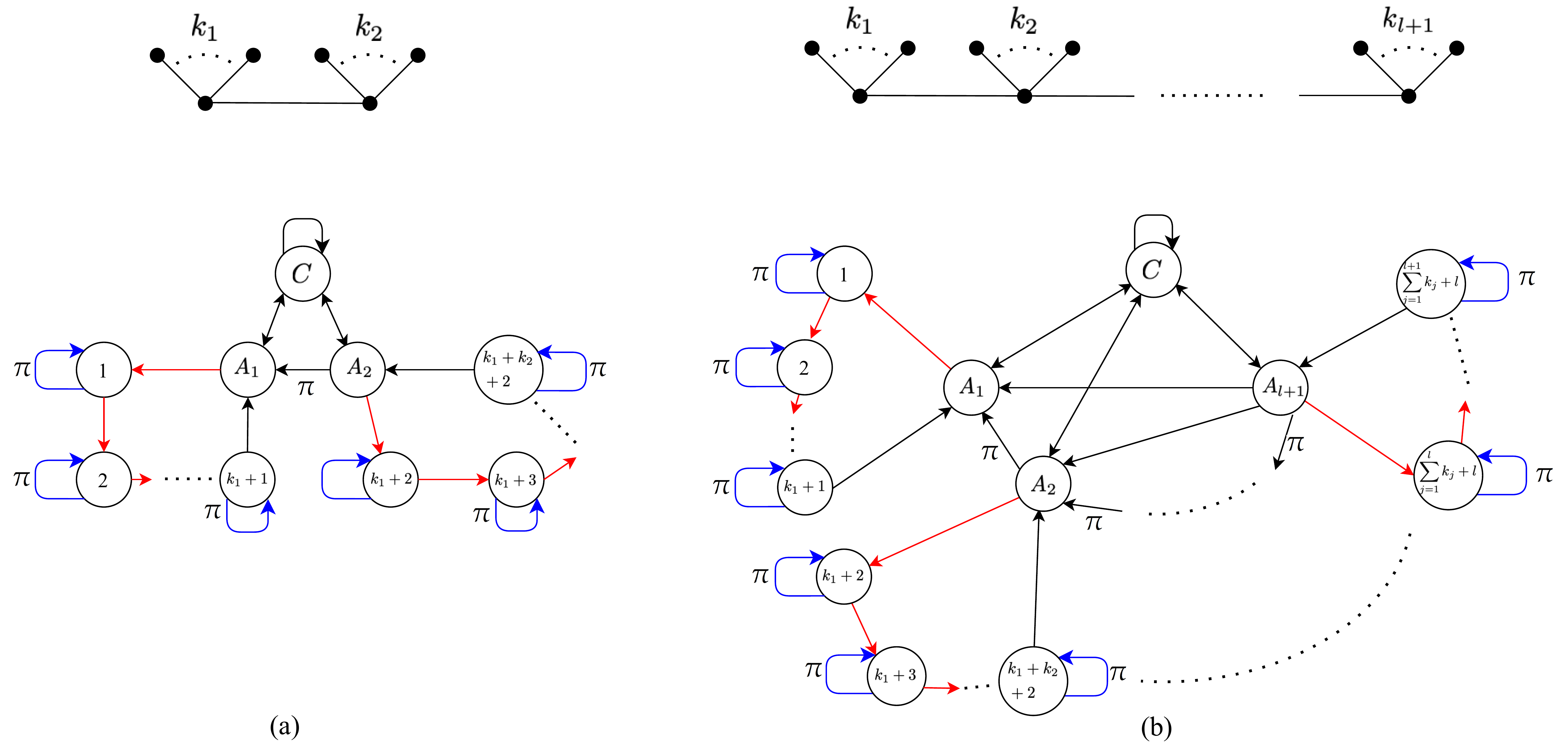}
        \caption{Sculpting bigraphs that generate CGSs. (a) A sculpting digraph for an arbitrary $(k_1+k_2+2)$-partite CGS of path length 1, directly generalizing the digraph~\eqref{Bell_di}. (b) A sculpting digraph for an arbitrary $\sum_{l}(k_l+1)$-partite caterpillar graph state of length $l$ based on $P^{(l)}$.}
        \label{fig:caterpillar_l}
    \end{figure*}    

We can use the path circuit digraphs $P^{(l)}$ to construct sculpting bigraphs that correspond to the sculpting operators that generate CGSs of length $l$. To achieve that, we use the one-to-one correspondence between the DCCs in a path circuit digraph and those in a sculpting digraph that replaces some of loops with \emph{primate circuit  digraphs}, which is an open graph of the following form:
   \begin{align}
\includegraphics[width=.25\textwidth]{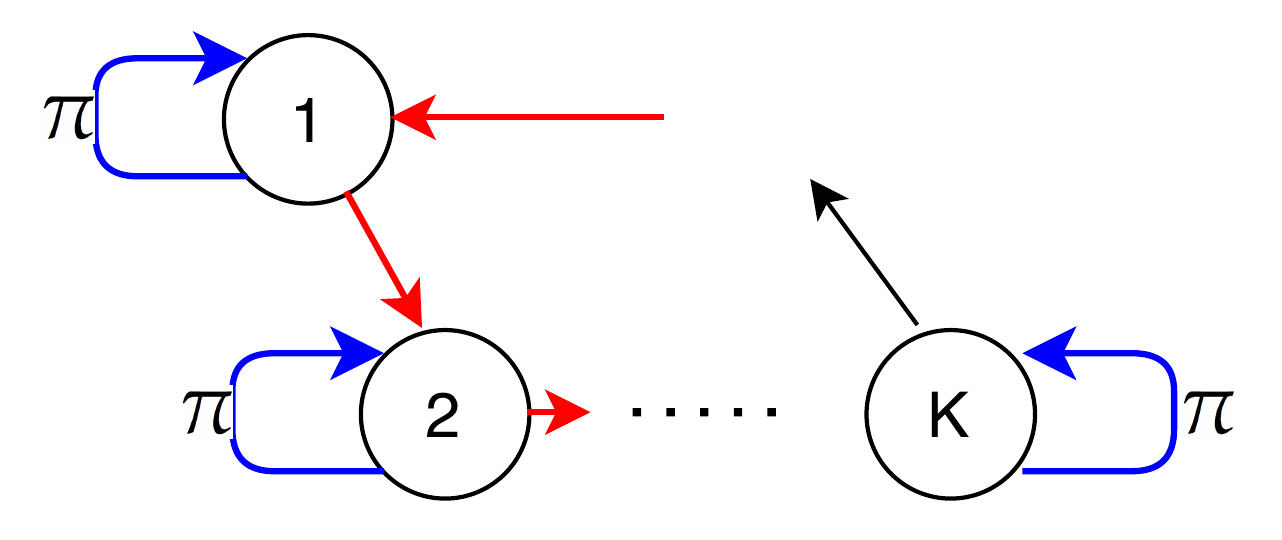} \nn 
   \end{align} where the red incoming edge to the circle $1$ and the black outgoing edge from the circle $K$ will be attached to $A_j$ ($j\in \{1,2,\cdots, l+1\}$). 
Then we can show that  $P^{(l)}$ and primate circuit digraphs can combine to generate arbitrary CGSs of length $l$ by replacing all the loops with primate circuit graphs. The proof is given in Appendix~\ref{pl_proof}.

 To demonstrate the concept, we begin with the simplest CGS, i.e., a path of length 1,
	\begin{align}\label{N=2}
		\begin{tikzpicture}[baseline={([yshift=-.5ex]current bounding box.center)}]
			\node[fill, vertex] (1) at (0,0) { };
			\node[fill, vertex] (2) at (1,0) { };
			\path[line width = 0.6pt] (1) edge  (2);
		\end{tikzpicture}
		=|00\> +|01\>+ |10\> - |11\>.
	\end{align} 
We can generate the above state with the following sculpting digraph, which is a combination of $P^{(1)}$ and two primate circuit  digraphs:
	\begin{align}\label{Bell_di}
		\begin{gathered}
			\includegraphics[width=5.5cm]{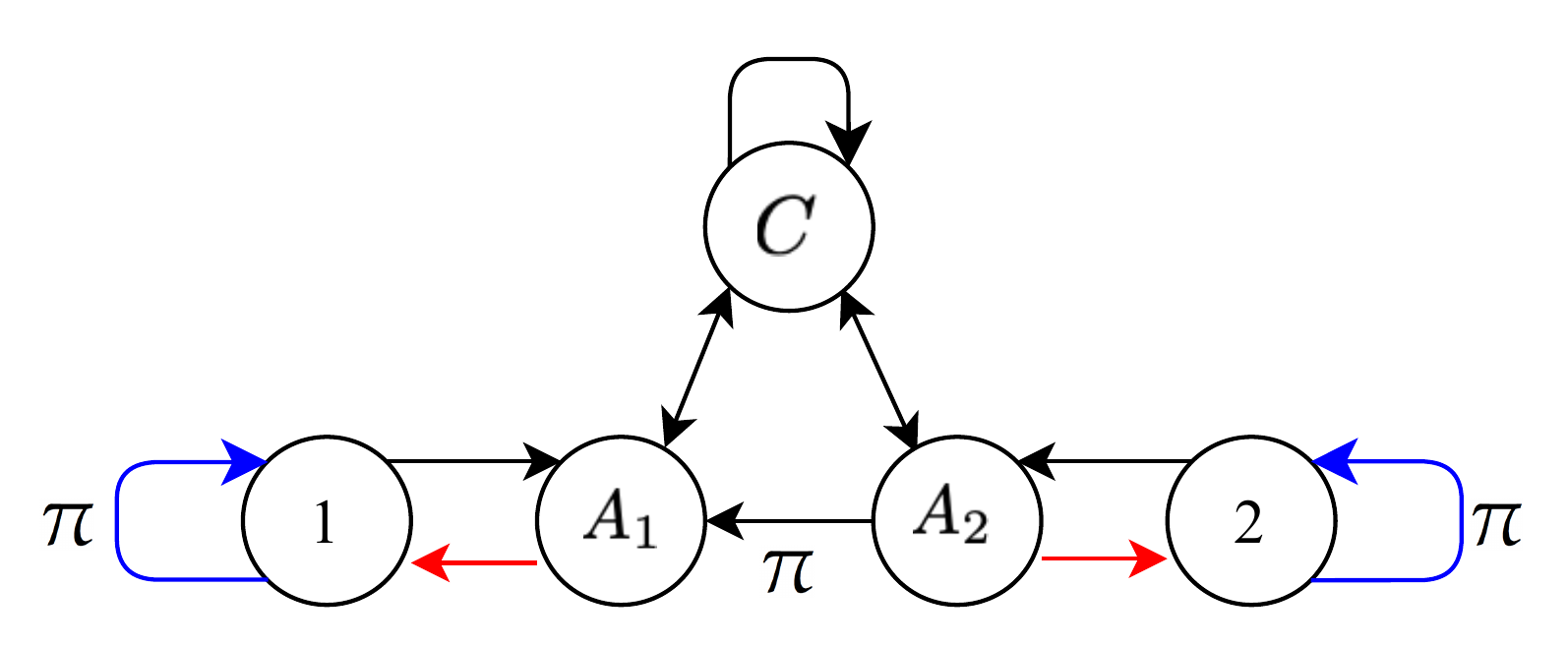}   
		\end{gathered}.
	\end{align} 
  The corresponding sculpting operator from Table~I is given by
\begin{align}\label{sculpting_2_op}
&\hat{A}_{2+P^{(1)}}\nn \\
&= (-\ha_{1,1} + \ha_{A_1,+})(-\ha_{2,1} + \ha_{A_2,+})(\ha_{1,0} + \ha_{C,+}) \nn \\
&~~\times(\ha_{2,0} - \ha_{A_1,+} + \ha_{C,+})(\ha_{A_1,+} + \ha_{A_2,+} + \ha_{C,+}).
\end{align}   Since \eqref{Bell_di} is an EPM digraph, only its DCCs  contribute to the final states. We can see that there are four DCCs
\begin{align}
     \includegraphics[width=.5\textwidth]{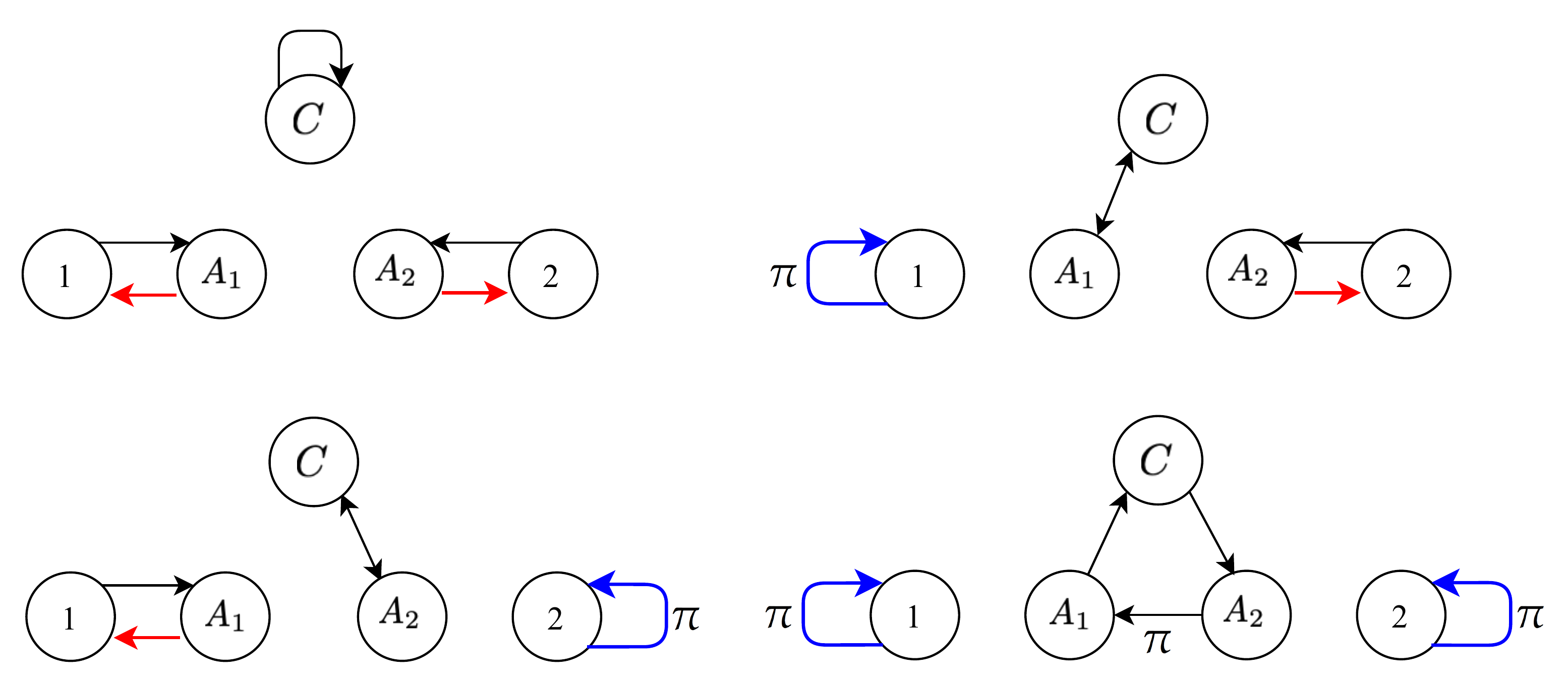},
\end{align}  which correspond to the following four terms:
\begin{align}
&\{ \ha_{1,0}\ha_{2,0} \ha_{A_1,+}\ha_{A_2,+}\ha_{C,+},- \ha_{1,0}\ha_{2,1} \ha_{A_1,+}\ha_{A_2,+}\ha_{C,+},\nn \\
&-\ha_{1,1}\ha_{2,0} \ha_{A_1,+}\ha_{A_2,+}\ha_{C,+},  -\ha_{1,1}\ha_{2,1} \ha_{A_1,+}\ha_{A_2,+}\ha_{C,+} \}.
 \end{align}
Applying the terms to the initial state
\begin{align}
 \Big(\prod_{j=1}^2\ha^\dagger_{j,+}\ha^\dagger_{j,-}\Big) \ha^\dagger_{A_1,+}\ha^\dagger_{A_2,+}\ha^\dagger_{C,+}|vac\>
\end{align} as defined in Eqs.~\eqref{initial} and \eqref{Anc},   
we obtain Eq.~\eqref{N=2}. 




	

The generalization of this digraph for an arbitrary CGS of path length $l$~\eqref{caterpillar_arb}
is straightforward. We start from $P^{(l)}$ and replace all the loops of $A_j$s ($j\in \{1,2,\cdots, l+1\}$) with primate circuit  digraphs of length $k_j+1$.  The entire sculpting digraph becomes an EPM digraph, as displayed in Fig.~\ref{fig:caterpillar_l}. 

\begin{figure*}
\centering 
    \includegraphics[width=.9\textwidth]{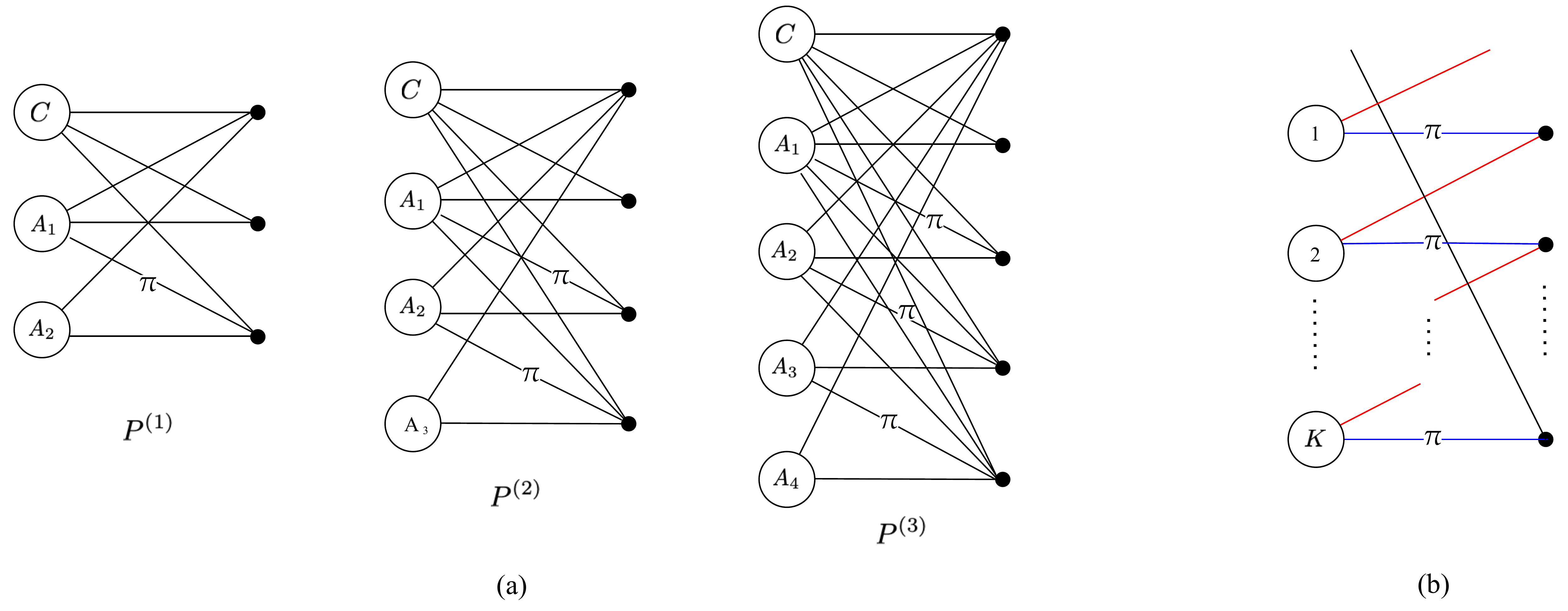}
    \caption{Elements that construct a sculpting bigraph for an arbitrary CGSs. (a) path circuit bigraphs $P^{(1)}$, $P^{(2)}$, and $P^{(3)}$ in $G_{ub}$. This is easily drawn from the directed graphs in Fig.~\ref{fig:central_path} using Table I. Note that all the horizontal lines correspond to the loops of path circuit digraphs, which will be replaced with the graph element (b), which is now called a primate circuit  bigraph.}
    \label{fig:central_path_bi}
\end{figure*}

    \begin{figure*}[t]\centering 
       \includegraphics[width=.8\textwidth]{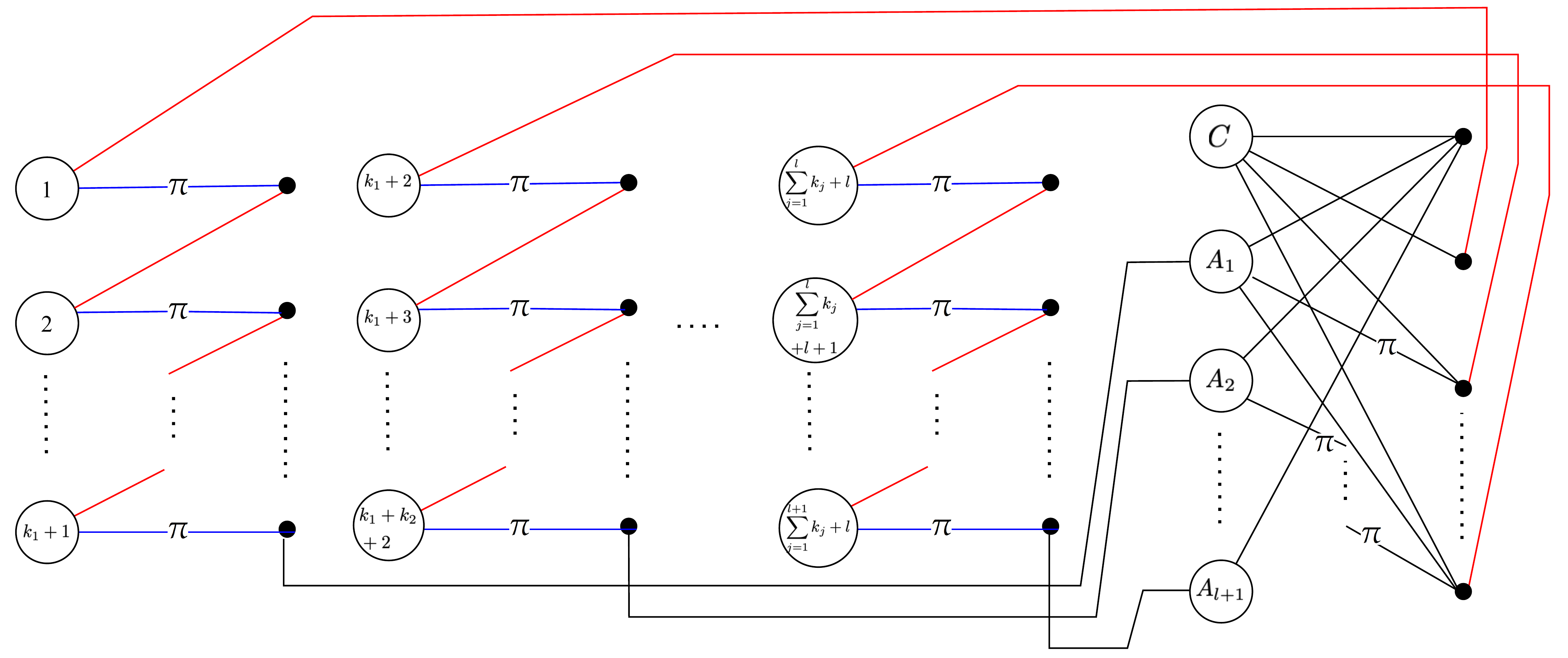}
        \caption{The sculpting bigraph that generates an arbitrary CGS of length $l$. The horizontal lines from $A_j$ $(j \in \{1,2,\cdots, l+1\})$ in $P^{(l)}$ are all replaced with primate circuit  bigraphs. For each $j$th primate circuit  bigraph, the open black edge is connected to $A_{j}$, and the open red edge to the dot next to $A_{j}$.} 
        \label{fig:caterpillar_l_bi}
    \end{figure*}

\section{Linear optical CGS circuits}\label{sec:circuits}

The sculpting digraph in Fig.~\ref{fig:caterpillar_l}(b) serves as a blueprint for the heralded linear optical circuit that generates arbitrary CGSs.
To design the circuit, we first transform the sculpting digraph to a corresponding sculpting bigraph in $G_{ub}$, which captures the structure of sculpting operators more intuitively. This transformation is directly achieved by comparing the second and third columns of Table~\ref{dict}.
Next, we use the translation rules from EPM bigraph elements to linear optical elements~\cite{chin2024heralded}. These rules were originally expressed in the polarization encoding but can be applied to dual-rail encoding with direct circuit modifications as in Fig.~7 of
Ref.~\cite{kang2026heralded}.
Then the translated optical components are connected following the corresponding graph structures.
Additionally, phase rotators with a phase shift $\pi$ are placed on the paths with edge weight $\pi$. 

To draw the sculpting bigraph, we decompose it into two parts, the path circuit bigraph $P^{(l)}$ and primate circuit  bigraphs following the sculpting digraph structure in Sec.~\ref{sec:sculpting_digraph} (see Fig.~\ref{fig:central_path_bi}).
As the loops of $A_j$ $(j~\in~\{1,2,\cdots, l+1\})$ in Fig.~\ref{fig:central_path} are replaced with primate circuit  digraphs for a complete sculpting digraph, the horizontal edges from $A_j$ $(j~\in~\{1,2,\cdots, l+1\})$ in Fig.~\ref{fig:central_path_bi} (a) are replaced with the primate circuit  bigraphs in Fig.~\ref{fig:central_path_bi} (b) to construct a complete sculpting bigraph. Fig.~\ref{fig:caterpillar_l_bi} displays the final bigraph form that corresponds to the sculpting digraph in Fig.~\ref{fig:caterpillar_l} (b).  
For each $j$th primate circuit  bigraph, the open black (red) edge is connected to $A_{j}$ (the dot next to $A_{j}$).

\begin{figure*}[t]
\centering 
    \includegraphics[width=1.\textwidth]{caterpillar_general_circuit}
    \caption{Linear optical circuit to generate an arbitrary CGS~\eqref{caterpillar_arb}. The circuit elements in the dotted box is from $P^{(l)}$, and those in the dashed boxes are from primate circuit  bigraphs (we now call them \emph{primate circuits}). Each solid box both in $P^{(l)}$ and primate circuits performs a single-photon subtraction. The black boxes with $\bar{p}$ are asymmetric $p$-partite splitter, in which the relative probability amplitudes are asymmetrically given as in Eq.~\eqref{app:asymmetric_amp}. The black boxes with $p$ are symmetric $p$-partite splitters that perform $p$-level discrete Fourier transforms. This circuit can also generate weighted graph states by relaxing the phase rotation angles.}  
    \label{fig:loc}
\end{figure*}

The linear optical circuit designed from the sculpting bigraph in Fig.~\ref{fig:caterpillar_l_bi} 
is drawn in  Fig.~\ref{fig:loc}. 
The circuit elements in the dotted box is from $P^{(l)}$, and those in the dashed boxes are from primate circuit  bigraphs (we now call them \emph{primate circuits}). Each solid box in $P^{(l)}$ and primate circuits performs a single-photon subtraction by postselecting events in which only one particle arrives at one of the detectors.

In each primate circuit , four rails in $K$th mode are denoted as $(K0,K0',K1,K1')$ as in the gray box of Fig.~\ref{fig:loc}, hence the initial state of the first primate circuit  is given by
\begin{align}
\prod^{k_1+1}_{K=1}\ha^\dagger_{K0}\ha^\dagger_{K1}|vac\>.
\end{align}

In $P^{(l)}$, each rail of a mode is denoted as $(L,s)$ (where $L$ and $s$ denote the mode and rail respectively), and the initial state of $P^{(l)}$ is given by
\begin{align}
    \ha^\dagger_{C0}\prod_{q=1}^{l+1}\had_{A_q 0}|vac\>.
\end{align}
$ $\\
The wires from $C,A_1,\cdots, A_{l+1}$ are connected to those subtractions as follows:
\begin{align}
 & \ha^\dagger_{c,0} \to \hb^\dagger_{c,0},~~\ha^\dagger_{c,j} \to \hb^\dagger_{j,0}, \nn \\
 & \ha^\dagger_{A_k,0} \to \hb^\dagger_{c,k},~~\ha^\dagger_{A_k,m} \to \hb^\dagger_{k+m,k}~~(1\le m \le l+1-k),\nn \\
 &\ha^\dagger_{A_k,l+2-k} \to \hb^\dagger_{\tilde{k},0}.     
\end{align}
The wires from primate circuits are connected to the subtractions so that the upper (lower) wire from primate circuit  $j$ is connected to $A_{j}$ ($\tilde{j}$).

Our scheme requires $2\sum_{j=1}^{l+1}k_j +3l+4$ photons ($2(\sum_{j=1}^{l+1}(k_j +1))$ photons in primate circuits and $l+2$ photons in $P^{(l)}$).  
The success probability for generating a CGS \eqref{caterpillar_arb} is given by  
\begin{align}\label{eq:success_prob}
    P_{l, \vec{k}}^{LQN} = \frac{1}{2^{2\sum_{j=1}^{l+1} k_j + 3l+3}}.
\end{align}
Appendix~\ref{app:suc_prob} provides a detailed explanation on the operation of this circuit with a rigorous success probability calculation \footnote{
It is worth mentioning that our scheme can be directly generalized for generating weighted graph states~\cite{sen2006quantum,anders2007variational}, in which the CZ gates are replaced with weighted CZ gates. In our scheme, it is achieved by relaxing the phase rotation angles $\pi$ to arbitrary angles $\phi_j$ for each $A_j$. 
}.

\paragraph*{Comparison with the fusion-based method.---}
A conventional way to generate the CGS is to use fusion gates~\cite{browne2005resource,varnava2006loss,varnava2008good}. The idea is to prepare for (l+1) GHZ states (star graph states), which are fused with Type-I fusion gates to generate the CGS.  The protocol is summarized as follows:
\begin{enumerate}
    \item Prepare for $(k_j+2)$-partite GHZ states for $j \in \{1,2,
    \cdots  , l\}$ and a $(k_{l+1}+1)$-partite GHZ states using the heralded scheme introduced in Ref.~\cite{chin2024heralded}, which  generates $N$-partite GHZ states with $2N$ photons and $1/2^{2N-1}$ of success probability. 
    \item Connect them using Type I fusion gates~\cite{browne2005resource} as follows:
\begin{align}
    \includegraphics[width=.4\textwidth]{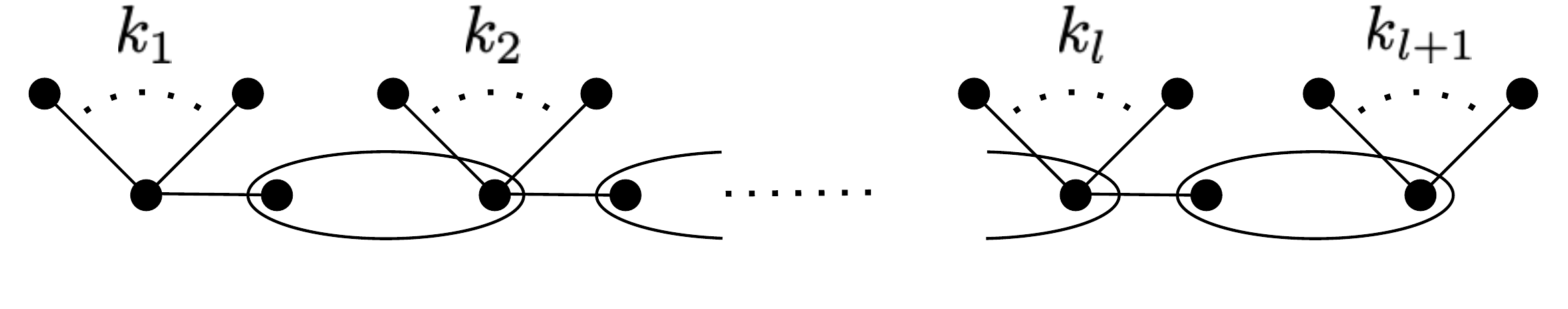} \nn 
\end{align} where each circle corresponds to a Type-I fusion gate.

\end{enumerate}
 After the successful fusion, the final state is given by
\begin{align}\label{caterpillar_arb2}
    \includegraphics[width=.45\textwidth]{caterpillar_arb.pdf},
\end{align} the CGS.

The required photon number of this protocol is $2\sum_{j=1}^{l+1}k_j+4l+2$ and the success probability is given by
\begin{align}
       P_{l, \vec{k}}^{Type-I} = \frac{1}{2^{2\sum_{j=1}^{l+1} k_j + 4l+1}}.
\end{align}
Therefore, comparing with Eq.~\eqref{eq:success_prob}, we can see that within heralded-optical implementations, the LQG-picture-based scheme is more efficient than the fusion-based scheme in both the photon number and success probability for $l \ge 3$. Indeed, the LQG-picture-based scheme saves $(l-2)$ photons with $2^{l-2}$ times higher success probability for $l\ge 3$.

In measurement-based quantum computing (MBQC),
we can use CGSs as resource states for generating arbitrary graph states with additional fusion gates as proposed in Refs.~\cite{hilaire2023near,lobl2025generating}.  Therefore, we can state that the cost reduction of resource-state generation with LQG framework directly lowers the overall resource overhead of the entire computation process.

\section{Concluding discussion and outlook}\label{sec:discussions}

We have proposed an efficient scheme for generating arbitrary caterpillar graph states (CGSs) using only single-photon sources, linear optical elements, and photon-number-resolving detectors. Guided by the linear quantum graph (LQG) picture, our approach employs flexible heralding measurements that enable scalable and resource-efficient state generation. The same framework can be extended to generalized CGSs and other graph families by adapting the underlying LQG structures.
A central insight is that fusion gates can be regarded as a special class of heralded boson subtractions. Both correspond to specific measurement configurations that create heralded entanglement through selective photon detection. The LQG picture unifies these processes, providing a systematic all-photonic route for reducing the preparation overhead of selected MBQC resource states.

Experimentally, the scheme is feasible with current photonic technology. It is useful for the efficient generation of arbitrarily high-dimensional cluster states in MBQC.  It also offers an optimal platform 
for testing local-operations-and-shared-randomness genuine multipartite nonlocality (LOSR-GMNL). Ref. [15] shows that all CGSs exhibit LOSR-GMNL, unlike linear graph states or GHZ states. Our optimized scheme provides a more practical circuit design than the fusion method, especially for current laboratory-scale experiments.

Quantum-emitter approaches~\cite{schwartz2016deterministic,russo2019generation,thomas2022efficient,shapourian2023modular,hilaire2023near,huet2025deterministic} provide an important alternative route to caterpillar and more general photonic graph states. In the ideal setting, sequential emission protocols can generate large states using one, or a small number of, coherently controlled emitters and may therefore require fewer independently prepared optical inputs than the present construction. 
These schemes, however, rely on efficient photon extraction, repeated coherent control, and sufficiently long matter-qubit coherence.  Errors in these sequential operations may accumulate as the generated state size increases~\cite{huet2025deterministic}.
For subsequent fusion-based processing, emitter-generated states also require integration with linear-optical networks, which may introduce additional experimental imperfections.
The present proposal avoids these matter-qubit and hybrid-integration requirements, but  at the cost of an increasing  interferometer size and an exponentially decreasing one-shot success probability. The two approaches therefore address different experimental regimes. A platform-specific comparison with emitter-based generation would require a common model of loss, operation time, coherence and control error.

Looking ahead, the LQG framework naturally generalizes to more complex graph-state families and could be extended to hybrid photonic–matter systems~\cite{moehring2007entanglement,ritter2012elementary,nemoto2014photonic}, adaptive feedforward architectures~\cite{scheel2006feed}, and integrated photonic circuits~\cite{obrien2007}. Such developments would further enhance efficiency and scalability, paving the way toward practical large-scale implementations of measurement-based quantum information processing.

	\section*{Acknowledgements}
	The authors are grateful to John Selby, Ana Belen Sainz, Yong-Su Kim, Marcin Karczewski, Paweł Cie\'{s}li\'{n}ski, and Waldemar K\l{}obus  for fruitful discussions. 
	This research was supported by
National Research Foundation of Korea (NRF, RS-2023-00245747) and the quantum computing technology development program of the National Research Foundation of Korea (NRF) funded by the Korean government (Ministry of Science and ICT, MSIT, No.2021M3H3A103657313).

 	\bibliographystyle{unsrt}
	\bibliography{graphtograph}

\begin{thebibliography}{10}

\bibitem{knill2001scheme}
Emanuel Knill, Raymond Laflamme, and Gerald~J Milburn.
\newblock A scheme for efficient quantum computation with linear optics.
\newblock {\em Nature}, 409(6816):46--52, 2001.
\newblock \url{https://doi.org/10.1038/35051009}.

\bibitem{kok2007linear}
Pieter Kok, William~J Munro, Kae Nemoto, Timothy~C Ralph, Jonathan~P Dowling,
  and Gerard~J Milburn.
\newblock Linear optical quantum computing with photonic qubits.
\newblock {\em Reviews of Modern Physics}, 79(1):135--174, 2007.
\newblock \url{https://doi.org/10.1103/revmodphys.79.135}.

\bibitem{obrien2007}
Jeremy~L. O'Brien.
\newblock Optical quantum computing.
\newblock {\em Science}, 318(5856):1567--1570, 2007.
\newblock \url{https://doi.org/10.1126/science.1142892}.

\bibitem{azuma2015all}
Koji Azuma, Kiyoshi Tamaki, and Hoi-Kwong Lo.
\newblock All-photonic quantum repeaters.
\newblock {\em Nature Communications}, 6(1):6787, 2015.
\newblock \url{https://doi.org/10.1038/ncomms7787}.

\bibitem{azuma2015allqkd}
Koji Azuma, Kiyoshi Tamaki, and William~J Munro.
\newblock All-photonic intercity quantum key distribution.
\newblock {\em Nature Communications}, 6(1):10171, 2015.
\newblock \url{https://doi.org/10.1038/ncomms10171 }.

\bibitem{raussendorf2001one}
Robert Raussendorf and Hans~J Briegel.
\newblock A one-way quantum computer.
\newblock {\em Physical Review Letters}, 86(22):5188, 2001.
\newblock \url{https://doi.org/10.1103/physrevlett.86.5188}.

\bibitem{raussendorf2003measurement}
Robert Raussendorf, Daniel~E Browne, and Hans~J Briegel.
\newblock Measurement-based quantum computation on cluster states.
\newblock {\em Physical Review A}, 68(2):022312, 2003.
\newblock \url{https://doi.org/10.1103/physreva.68.022312}.

\bibitem{hein2004multiparty}
Marc Hein, Jens Eisert, and Hans~J Briegel.
\newblock Multiparty entanglement in graph states.
\newblock {\em Physical Review A}, 69(6):062311, 2004.
\newblock \url{https://doi.org/10.1103/physreva.69.062311}.

\bibitem{briegel2009measurement}
Hans~J Briegel, David~E Browne, Wolfgang D{\"u}r, Robert Raussendorf, and
  Maarten Van~den Nest.
\newblock Measurement-based quantum computation.
\newblock {\em Nature Physics}, 5(1):19--26, 2009.
\newblock \url{https://doi.org/10.1038/nphys1157}.

\bibitem{van2013universal}
Maarten Van~den Nest.
\newblock Universal quantum computation with little entanglement.
\newblock {\em Physical Review Letters}, 110(6):060504, 2013.
\newblock \url{https://doi.org/10.1103/physrevlett.110.060504 }.

\bibitem{hein2006entanglement}
Marc Hein, Wolfgang D{\"u}r, Jens Eisert, Robert Raussendorf, M~Nest, and H-J
  Briegel.
\newblock Entanglement in graph states and its applications.
\newblock {\em arXiv preprint quant-ph/0602096}, 2006.
\newblock \url{https://doi.org/10.48550/arXiv.quant-ph/0602096}.

\bibitem{schlingemann}
D.~Schlingemann and R.~F. Werner.
\newblock Quantum error-correcting codes associated with graphs.
\newblock {\em Phys. Rev. A}, 65:012308, Dec 2001.
\newblock \url{https://doi.org/10.1103/PhysRevA.65.012308}.

\bibitem{markham2008}
Damian Markham and Barry~C. Sanders.
\newblock Graph states for quantum secret sharing.
\newblock {\em Phys. Rev. A}, 78:042309, Oct 2008.
\newblock \url{https://doi.org/10.1103/PhysRevA.78.042309}.

\bibitem{bell2014experimental}
BA~Bell, Damian Markham, DA~Herrera-Mart{\'\i}, Anne Marin, WJ~Wadsworth,
  JG~Rarity, and MS~Tame.
\newblock Experimental demonstration of graph-state quantum secret sharing.
\newblock {\em Nature Communications}, 5(1):1--12, 2014.
\newblock \url{https://doi.org/10.1038/ncomms6480}.

\bibitem{coiteux2025genuinely}
Xavier Coiteux-Roy, Owidiusz Makuta, Fionnuala Curran, Remigiusz Augusiak, and
  Marc-Olivier Renou.
\newblock The genuinely multipartite nonlocality of graph states is
  model-dependent.
\newblock {\em npj Quantum Information}, 11(1):1--6, 2025.
\newblock \url{https://doi.org/10.1038/s41534-025-01024-x}.

\bibitem{browne2005resource}
Daniel~E Browne and Terry Rudolph.
\newblock Resource-efficient linear optical quantum computation.
\newblock {\em Physical Review Letters}, 95(1):010501, 2005.
\newblock \url{https://doi.org/10.1103/physrevlett.95.010501}.

\bibitem{varnava2006loss}
Michael Varnava, Daniel~E Browne, and Terry Rudolph.
\newblock Loss tolerance in one-way quantum computation via counterfactual
  error correction.
\newblock {\em Physical Review Letters}, 97(12):120501, 2006.
\newblock \url{https://doi.org/10.1103/physrevlett.97.120501}.

\bibitem{varnava2008good}
Michael Varnava, Daniel~E Browne, and Terry Rudolph.
\newblock How good must single photon sources and detectors be for efficient
  linear optical quantum computation?
\newblock {\em Physical Review Letters}, 100(6):060502, 2008.
\newblock \url{https://doi.org/10.1103/physrevlett.100.060502}.

\bibitem{li2015resource}
Ying Li, Peter~C Humphreys, Gabriel~J Mendoza, and Simon~C Benjamin.
\newblock Resource costs for fault-tolerant linear optical quantum computing.
\newblock {\em Physical Review X}, 5(4):041007, 2015.
\newblock \url{https://doi.org/10.1103/physrevx.5.041007}.

\bibitem{bartolucci2023fusion}
Sara Bartolucci, Patrick Birchall, Hector Bombin, Hugo Cable, Chris Dawson,
  Mercedes Gimeno-Segovia, Eric Johnston, Konrad Kieling, Naomi Nickerson,
  Mihir Pant, et~al.
\newblock Fusion-based quantum computation.
\newblock {\em Nature Communications}, 14(1):912, 2023.
\newblock \url{https://doi.org/10.1038/s41467-023-36493-1}.

\bibitem{lee2023graph}
Seok-Hyung Lee and Hyunseok Jeong.
\newblock Graph-theoretical optimization of fusion-based graph state
  generation.
\newblock {\em Quantum}, 7:1212, 2023.
\newblock \url{https://doi.org/10.22331/q-2023-12-20-1212}.

\bibitem{chin2021graph}
Seungbeom Chin, Yong-Su Kim, and Sangmin Lee.
\newblock Graph picture of linear quantum networks and entanglement.
\newblock {\em Quantum}, 5:611, 2021.
\newblock \url{https://doi.org/10.22331/q-2021-12-23-611}.

\bibitem{chin2024shortcut}
Seungbeom Chin, Yong-Su Kim, and Marcin Karczewski.
\newblock Shortcut to multipartite entanglement generation: A graph approach to
  boson subtractions.
\newblock {\em npj Quantum Information}, 10(1):67, 2024.
\newblock \url{https://doi.org/10.1038/s41534-024-00845-6}.

\bibitem{chin2024heralded}
Seungbeom Chin, Marcin Karczewski, and Yong-Su Kim.
\newblock Heralded optical entanglement generation via the graph picture of
  linear quantum networks.
\newblock {\em Quantum}, 8:1572, 2024.
\newblock \url{https://doi.org/10.22331/q-2024-12-18-1572}.

\bibitem{chin2024exponentially}
Seungbeom Chin, Junghee Ryu, and Yong-Su Kim.
\newblock Exponentially enhanced scheme for the heralded qudit
  greenberger-horne-zeilinger state in linear optics.
\newblock {\em Physical Review Letters}, 133(25):253601, 2024.
\newblock \url{https://doi.org/10.1103/physrevlett.133.253601}.

\bibitem{hilaire2023near}
Paul Hilaire, Leonid Vidro, Hagai~S Eisenberg, and Sophia~E Economou.
\newblock Near-deterministic hybrid generation of arbitrary photonic graph
  states using a single quantum emitter and linear optics.
\newblock {\em Quantum}, 7:992, 2023.
\newblock \url{https://doi.org/10.22331/q-2023-04-27-992}.

\bibitem{pettersson2025deterministic}
Love~A Pettersson, Anders~S S{\o}rensen, and Stefano Paesani.
\newblock Deterministic generation of concatenated graph codes from quantum
  emitters.
\newblock {\em PRX Quantum}, 6(1):010305, 2025.
\newblock \url{https://doi.org/10.1103/prxquantum.6.010305}.

\bibitem{kang2026heralded}
Minhyeok Kang, Jaehee Kim, William~J Munro, Seungbeom Chin, and Joonsuk Huh.
\newblock Heralded linear optical generation of dicke states.
\newblock {\em New Journal of Physics}, 28(5):054501, 2026.
\newblock \url{https://doi.org/10.1088/1367-2630/ae6135}.

\bibitem{sen2006quantum}
Aditi Sen, Ujjwal Sen, Veronica Ahufinger, Hans~J Briegel, Anna Sanpera, and
  Maciej Lewenstein.
\newblock Quantum-information processing in disordered and complex quantum
  systems.
\newblock {\em Physical Review A—Atomic, Molecular, and Optical Physics},
  74(6):062309, 2006.
\newblock \url{https://doi.org/10.1103/physreva.74.062309}.

\bibitem{anders2007variational}
Simon Anders, Hans~J Briegel, and Wolfgang D{\"u}r.
\newblock A variational method based on weighted graph states.
\newblock {\em New Journal of Physics}, 9(10):361, 2007.
\newblock \url{https://doi.org/10.1088/1367-2630/9/10/361 }.

\bibitem{lobl2025generating}
Matthias~C L{\"o}bl, Love~A Pettersson, Andrew Jena, Luca Dellantonio, Stefano
  Paesani, and Anders~S S{\o}rensen.
\newblock Generating graph states with a single quantum emitter and the minimum
  number of fusions.
\newblock {\em Physical Review A}, 111(5):052604, 2025.
\newblock \url{ https://doi.org/10.1103/physreva.111.052604}.

\bibitem{schwartz2016deterministic}
Ido Schwartz, Dan Cogan, Emma~R Schmidgall, Yaroslav Don, Liron Gantz, Oded
  Kenneth, Netanel~H Lindner, and David Gershoni.
\newblock Deterministic generation of a cluster state of entangled photons.
\newblock {\em Science}, 354(6311):434--437, 2016.
\newblock \url{https://doi.org/10.1126/science.aah4758 }.

\bibitem{russo2019generation}
Antonio Russo, Edwin Barnes, and Sophia~E Economou.
\newblock Generation of arbitrary all-photonic graph states from quantum
  emitters.
\newblock {\em New Journal of Physics}, 21(5):055002, 2019.
\newblock \url{https://doi.org/10.1088/1367-2630/ab193d}.

\bibitem{thomas2022efficient}
Philip Thomas, Leonardo Ruscio, Olivier Morin, and Gerhard Rempe.
\newblock Efficient generation of entangled multiphoton graph states from a
  single atom.
\newblock {\em Nature}, 608(7924):677--681, 2022.
\newblock \url{https://doi.org/10.1038/s41586-022-04987-5}.

\bibitem{shapourian2023modular}
Hassan Shapourian and Alireza Shabani.
\newblock Modular architectures to deterministically generate graph states.
\newblock {\em Quantum}, 7:935, 2023.
\newblock \url{https://doi.org/10.22331/q-2023-03-02-935}.

\bibitem{huet2025deterministic}
H~Huet, PR~Ramesh, SC~Wein, N~Coste, P~Hilaire, N~Somaschi, M~Morassi,
  A~Lema{\^\i}tre, Isabelle Sagnes, MF~Doty, et~al.
\newblock Deterministic and reconfigurable graph state generation with a single
  solid-state quantum emitter.
\newblock {\em Nature communications}, 16(1):4337, 2025.
\newblock \url{https://doi.org/10.1038/s41467-025-59693-3}.

\bibitem{moehring2007entanglement}
David~L Moehring, Peter Maunz, Steve Olmschenk, Kelly~C Younge, Dzmitry~N
  Matsukevich, L-M Duan, and Christopher Monroe.
\newblock Entanglement of single-atom quantum bits at a distance.
\newblock {\em Nature}, 449(7158):68--71, 2007.
\newblock \url{https://doi.org/10.1038/nature06118}.

\bibitem{ritter2012elementary}
Stephan Ritter, Christian N{\"o}lleke, Carolin Hahn, Andreas Reiserer, Andreas
  Neuzner, Manuel Uphoff, Martin M{\"u}cke, Eden Figueroa, Joerg Bochmann, and
  Gerhard Rempe.
\newblock An elementary quantum network of single atoms in optical cavities.
\newblock {\em Nature}, 484(7393):195--200, 2012.
\newblock \url{https://doi.org/10.1038/nature11023}.

\bibitem{nemoto2014photonic}
Kae Nemoto, Michael Trupke, Simon~J Devitt, Ashley~M Stephens, Burkhard
  Scharfenberger, Kathrin Buczak, Tobias N{\"o}bauer, Mark~S Everitt, J{\"o}rg
  Schmiedmayer, and William~J Munro.
\newblock Photonic architecture for scalable quantum information processing in
  diamond.
\newblock {\em Physical Review X}, 4(3):031022, 2014.
\newblock \url{https://doi.org/10.1103/physrevx.4.031022}.

\bibitem{scheel2006feed}
Stefan Scheel, William~J Munro, Jens Eisert, Kae Nemoto, and Pieter Kok.
\newblock Feed-forward and its role in conditional linear optical quantum
  dynamics.
\newblock {\em Physical Review A—Atomic, Molecular, and Optical Physics},
  73(3):034301, 2006.
\newblock \url{https://doi.org/10.1103/physreva.73.034301}.

\bibitem{dakna1997generating}
Mohammed Dakna, Tiemo Anhut, T~Opatrn{\`y}, Ludwig Kn{\"o}ll, and D-G Welsch.
\newblock Generating schr{\"o}dinger-cat-like states by means of conditional
  measurements on a beam splitter.
\newblock {\em Physical Review A}, 55(4):3184, 1997.
\newblock \url{https://doi.org/10.1103/physreva.55.3184}.

\bibitem{ourjoumtsev2006generating}
Alexei Ourjoumtsev, Rosa Tualle-Brouri, Julien Laurat, and Philippe Grangier.
\newblock Generating optical schrodinger kittens for quantum information
  processing.
\newblock {\em Science}, 312(5770):83--86, 2006.
\newblock \url{https://doi.org/10.1126/science.1122858}.

\bibitem{zavatta2008subtracting}
A~Zavatta, V~Parigi, MS~Kim, and M~Bellini.
\newblock Subtracting photons from arbitrary light fields: experimental test of
  coherent state invariance by single-photon annihilation.
\newblock {\em New Journal of Physics}, 10(12):123006, 2008.
\newblock \url{https://doi.org/10.1088/1367-2630/10/12/123006}.

\bibitem{averchenko2016multimode}
V~Averchenko, C~Jacquard, V~Thiel, C~Fabre, and N~Treps.
\newblock Multimode theory of single-photon subtraction.
\newblock {\em New Journal of Physics}, 18(8):083042, 2016.
\newblock \url{https://doi.org/10.1088/1367-2630/18/8/083042}.

\bibitem{bart2021creation}
Sara Bartolucci, Patrick~M Birchall, Mercedes Gimeno-Segovia, Eric Johnston,
  Konrad Kieling, Mihir Pant, Terry Rudolph, Jake Smith, Chris Sparrow, and
  Mihai~D Vidrighin.
\newblock Creation of entangled photonic states using linear optics.
\newblock {\em arXiv preprint arXiv:2106.13825}, 2021.
\newblock \url{https://doi.org/10.48550/arXiv.2106.13825}.

\bibitem{reck1994experimental}
Michael Reck, Anton Zeilinger, Herbert~J Bernstein, and Philip Bertani.
\newblock Experimental realization of any discrete unitary operator.
\newblock {\em Physical Review Letters}, 73(1):58, 1994.
\newblock \url{https://doi.org/10.1103/physrevlett.73.58}.

\bibitem{lim2005generalized}
Yuan~Liang Lim and Almut Beige.
\newblock Generalized {H}ong--{O}u--{M}andel experiments with bosons and
  fermions.
\newblock {\em New Journal of Physics}, 7(1):155, 2005.
\newblock \url{https://doi.org/10.1088/1367-2630/7/1/155}.

\end{thebibliography}
	
\onecolumn\newpage
\appendix
	
	\section{Graph theory preliminaries}
\label{glossary}

\begin{figure}[b]
\centering 
    \includegraphics[width=.5\textwidth]{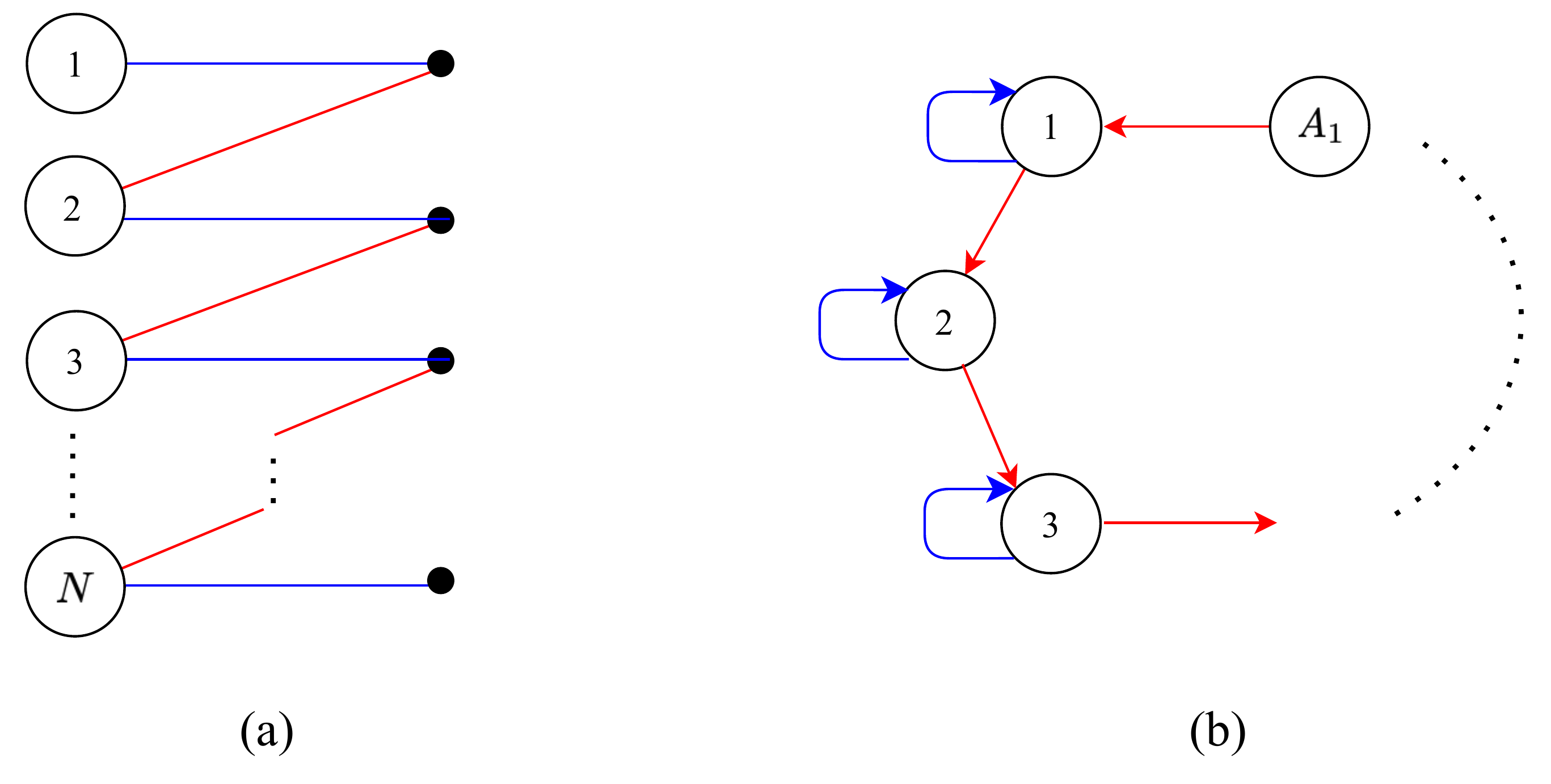}
    \caption{Sculpting operators for generating $N$-partite GHZ states represented as (a) sculpting bigraphs and (b) sculpting digraphs~\cite{chin2024shortcut}. With Table I, they correspond to a sculpting operators $\hat{A}_N = \frac{1}{\sqrt{2^N}}  (\ha_{1,0} + \ha_{2,1})(\ha_{2,0} + \ha_{3,1})\cdots (\ha_{N-1,0} + \ha_{N,1})(\ha_{N,0} + \ha_{1,1})$.  We choose a \emph{simplified notation} here, which is valid when  all the edges connected to the same dot has the same real value of edge weights. In the simplified notation, omission of edge weights  implies that the edge weight is real and positive. An edge weight $\pi$ implies that the edge weight is real and negative.}
    \label{fig:GHZ}
\end{figure}       
    
	\paragraph*{Graph---} A \textbf{graph} $G=(V,E)$ is a set of vertices $V$ whose elements are connected by edges $(\in E)$. Each edge can have some numerical values, \textbf{weights}. 	
	\paragraph*{Undirected and directed graphs---.} Edges in a graph can have directions. \textbf{undirected graphs} (\textbf{directed graphs}, digraphs $G_d$) have undirected (directed) edges. For an undirected graph $G=(V,E)$ with $V= \{v_1,v_2,\cdots, v_N\}$, we denote an edge ($\in E$) that connects $v_i$ and $v_j$ as $(v_i,v_j)$.  
	For a $G_d=(W,F)$ with $W=\{w_1,w_2,\cdots, w_N\}$, we denote an edge ($\in F$) that connects $w_i$ and $w_j$ as $(w_i\to w_j)$.  	
A digraph $G_d$ is \textbf{strongly connected} if every vertex has a path to every other vertex.
	
	\paragraph*{Bipartite graph and perfect matching---}  A \textbf{bipartite graph} (bigraph, $G_b=(U\cup V, E)$) has two sets of vertices, $U$ and $V$,  and its edges do not connect two vertices in the same set (on the contrary, when a graph has one set of vertices, it is called a \textbf{unipartite graph} (unigraph)). A \textbf{balanced bigraph}, denoted as $G_{bb}$, is a bigraph with $|U|=|V|$. All the bigraphs that we consider in this work are balanced bigraphs, e.g., Fig.~\ref{fig:GHZ}. A balanced bigraph can have  \textbf{perfect matchings} (PMs). A PM of $G_{bb}$ is a subgraph in which every vertex in $U$ is adjacent to only one edge in $V$.
	For example, the GHZ sculpting bigraph in Fig. ~\ref{fig:GHZ} (a) has two PMs,
	\begin{align}
		\begin{gathered}   \includegraphics[width=6cm]{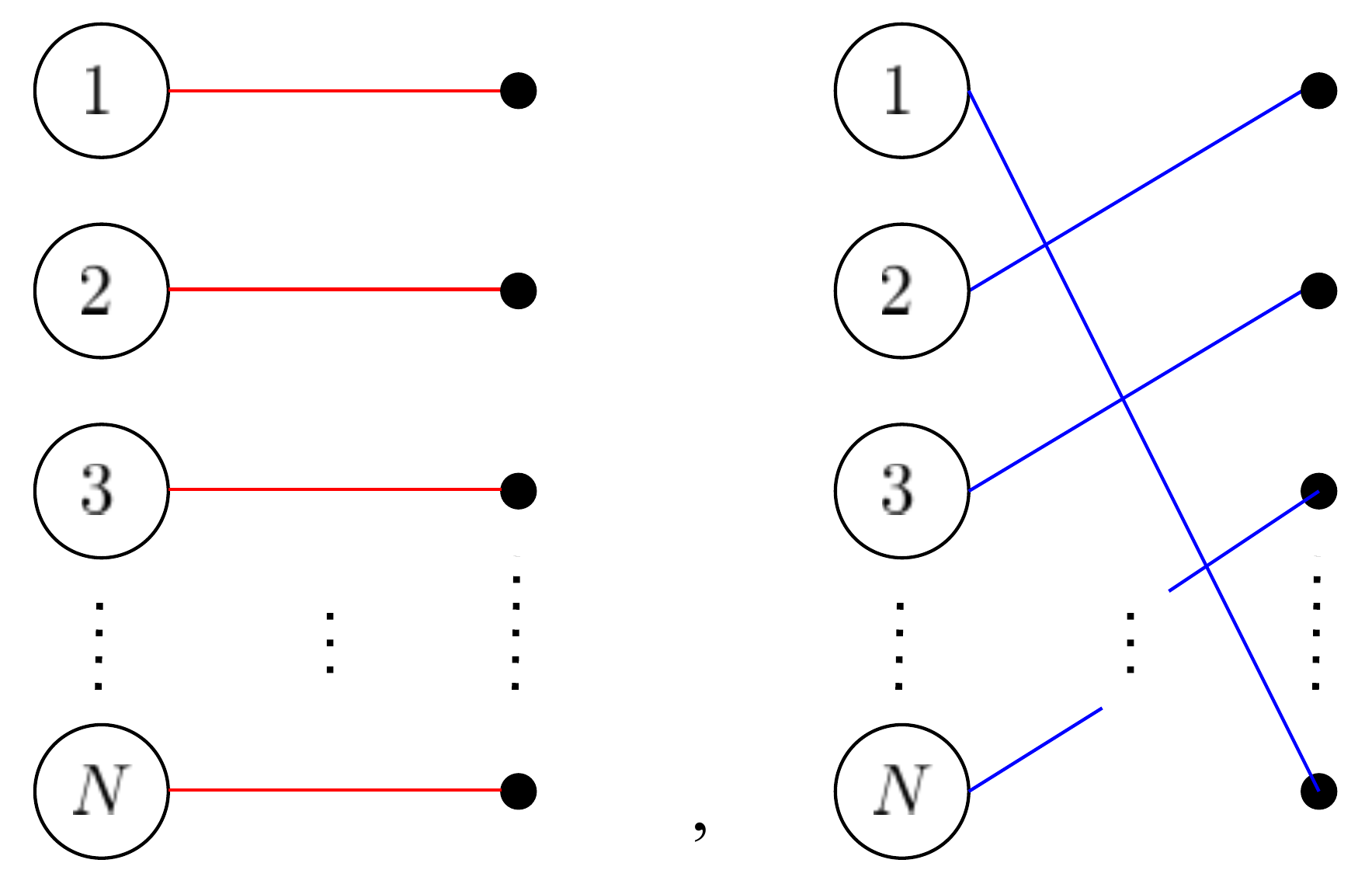}
		\end{gathered} 
	\end{align} 
	which correspond to the two terms of the GHZ state. 
	
	\paragraph*{From bigraphs to digraphs---} We can draw a \textbf{unipartite digraph} that corresponds to a bigraph using the following definition:
\begin{definition}\label{def:bi_to_di}
    (From bigraph to digraph)
    For a balanced bigraph $G_{ub} = (U\cup V, E)$, its digraph counterpart $G_{du}= (W,F)$ is obtained by mapping two vertices $v_i$ in $U$ and $v_{i'}$ in $V$ into a vertex $w_i$ in $W$, and a undirected edge $(v_i,v_{j'})$ ($\in E$) to a directed edge $(w_i \leftarrow  w_{j'})$ ($\in F$). 
\end{definition}
In our representation, labeled and unlabeled vertices are in  $U$ and $V$ respectively as explained in Table~\ref{dict}. 
    Then the digraph in Fig.~\ref{fig:GHZ} (b) is drawn from the bigraph in  Fig.~\ref{fig:GHZ} (a). 
    A PM in the bigraph correspond to a \textbf{disjoint cycle cover (DCC)}, in which every vertex has one incoming edge and one outgoing edge. For the GHZ sculpting digraph~\ref{fig:GHZ} (b), we can find two DCCs:
	\begin{align}
		\begin{gathered}   \includegraphics[width=.5\textwidth]{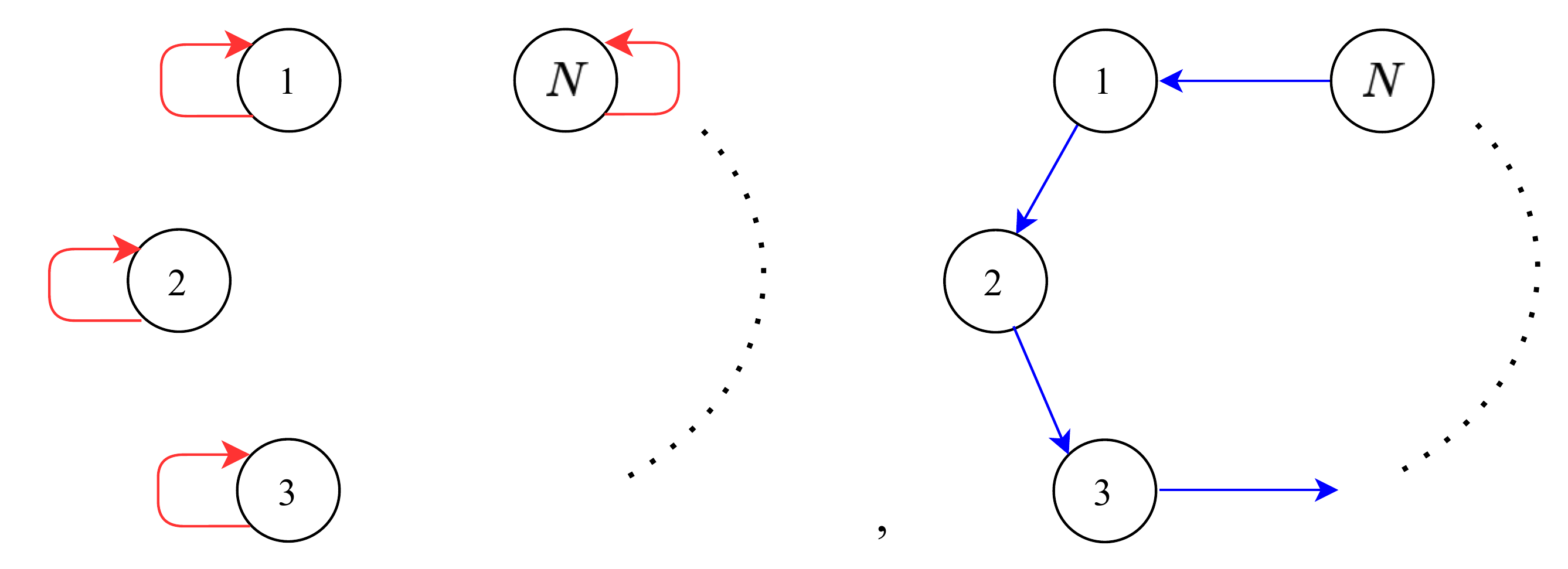}
		\end{gathered} 
	\end{align}

\section{Sculpting protocol} 
\label{sculpting}

The idea of generating entanglement through subtracting particles is originated from researchs on continious-variable photonic systems~\cite{dakna1997generating,ourjoumtsev2006generating,zavatta2008subtracting,averchenko2016multimode}.
In our setup for the sculpting protocol~\cite{chin2024shortcut}, 
each boson has an $N$-dimensional spatial state and a two-dimensional internal state, hence creation and annihilation operators are expressed as $\ha_{j,s}^\dagger$ and $\ha_{j,s}$ ($j\in \{1,2,\cdots, N\}$, $s \in \{+,-\}$) respectively.
The sculpting protocol consists of three steps for generating multipartite entanglement: 
\begin{enumerate}
    \item Initial state preparation: In the initial state,
 each mode has two bosons in the opposite internal states with each other: 
	\begin{align}\label{initial}
		|Sym_{N}\> \equiv \ha^\dagger_{1,+}\ha^\dagger_{1,-} \ha^\dagger_{2,+}\ha^\dagger_{2,-} \cdots 
		\ha^\dagger_{N,+}\ha^\dagger_{N,-}|vac\>  
		= \prod_{j=1}^N(\ha^\dagger_{j,+}\ha^\dagger_{j,-})|vac\>. 
	\end{align} 
 
Some sculpting schemes require a $K$-mode ancillary systems $(K \geq 0)$ in which each mode  has one particle of the same internal state (we choose $+$ here), i.e., 
 \begin{align}\label{Anc}
		|Anc_{K}\> \equiv \ha^\dagger_{N+1,+} \ha^\dagger_{N+2,+} \cdots 
		\ha^\dagger_{N+K,+}|vac\>  = \prod_{j=1}^K(\ha^\dagger_{N+j,+})|vac\>. 
	\end{align} 
Then the total initial state becomes $|Sym_{N}\>|Anc_{K}\>$ with $N+K$ spatial modes.  
\item Operation: We apply an $N$ single-boson annihilation operator (\emph{sculpting operator}), which is written as 
\begin{align}\label{annihilation}
		\prod_{l=1}^{N+K}\sum_{j=1}^{N+K}(k^{(l)}_{j,+}\ha_{j,+} + k^{(l)}_{j,-}\ha_{j,-} ) 
		\equiv \prod_{l=1}^{N+K}\hat{A}^{(l)} \equiv \hat{A}_{N+K} \quad (k^{(l)}_{j,s} \in \mathbb{C}~\textrm{and}~  \sum_{j,s}|k^{(l)}_{j,s}|^2 =1),
	\end{align}
 where $\hat{A}^{(l)}$ is a single-boson subtraction operator. 

The sculpting operator $\hat{A}_N$ must satisfy the \emph{no-bunching condition}, by which the final total state must be a superposition of states with one particle in each mode. From the viewpoint of heralded schemes, the remaining boson in the mode contains the qubit information, and the subtracted one plays the role of $heralding$.

 \item  Final state: We now obtain the final state  
	\begin{align}\label{final_state}
		|\P\>_{fin} = \hat{A}_{N+K}|Sym_{N}\>|Anc_{K}\>. 
	\end{align}
 We need to verify what type of entanglement $|\P\>_{fin}$ carries.  
\end{enumerate}



	\section{Caterpillar graph states (CGSs)}\label{CGS}
    
A controlled-Z (CZ) gate $U^{Z}_{jk}$ on $j$th and $k$th qubits is defined as
	\begin{align}
		U^{Z}_{jk}|r\>_j|s\>_k =(-1)^{rs}|r\>_j|s\>_k~~~(r,s\in \{0,1\})
	\end{align} 
	If a multi-qubit quantum state is a graph state, then qubits in the $+$ state correspond to vertices and CZ interactions between pairs of the qubits correspond to edges. More specifically, the graph state for a graph $G=(V,E)$ with $|V|=N$ is defined as
	\begin{align}
		|G\> = \prod_{(j,k)\in E}  U^{Z}_{jk}|\underbrace{++\cdots +}_{N}\> 
	\end{align}
	
	A caterpillar graph is a tree graph (a graph without a closed path) in which every vertex is on the path or one edge away from the path, e.g.,
	\begin{align}
		\includegraphics[width=7cm]{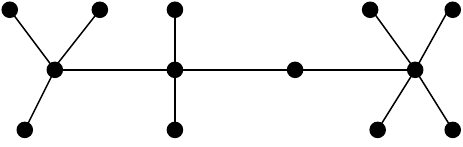}
	\end{align} is a caterpillar graph with a path of length 4. And a CGS is a graph state whose qubits are connected by CZ gates following the structure of a caterpillar graph. For example, three star graph states  (each has $K$, $L$, and $N-K$ qubits) whose centers are connected with path  length 2 is given by
	\begin{align}\label{eq:l=2}
		&\includegraphics[height=2cm]{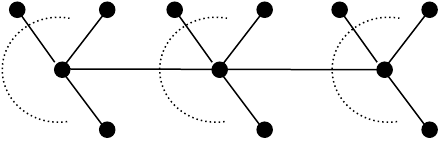} \nn \\
		&= |\underbrace{++\cdots +0}_{K}\>|\underbrace{++\cdots +0}_{L}\>|\underbrace{++\cdots +0}_{N-K-L}\> + |\underbrace{--\cdots -1}_{K}\>|\underbrace{++\cdots +0}_{L}\>|\underbrace{++\cdots +0}_{N-K-L}\>  \nn \\
		&~~+|\underbrace{++\cdots +0}_{K}\>|\underbrace{--\cdots -1}_{L}\>|\underbrace{++\cdots +0}_{N-K-L}\> - |\underbrace{--\cdots -1}_{K}\>|\underbrace{--\cdots -1}_{L}\>|\underbrace{++\cdots +0}_{N-K-L}\> \nn \\ 
		&~~+ |\underbrace{++\cdots +0}_{K}\>|\underbrace{++\cdots +0}_{L}\>|\underbrace{--\cdots -1}_{N-K-L}\>+ |\underbrace{--\cdots -1}_{K}\>|\underbrace{++\cdots +0}_{L}\>|\underbrace{--\cdots -1}_{N-K-L}\> \nn \\
		&~~- |\underbrace{++\cdots +0}_{K}\>|\underbrace{--\cdots -1}_{L}\> |\underbrace{--\cdots -1}_{N-K-L}\> + |\underbrace{--\cdots -1}_{K}\>|\underbrace{--\cdots -1}_{L}\> |\underbrace{--\cdots -1}_{N-K-L}\>.
	\end{align}

\section{Proof that the path circuit digraphs $P^{(l)}$ generates arbitrary CGS of length $l$}\label{pl_proof}
	
In this section, we show that any CGS of lenghth $l$ can be generated  by replacing the loops of $P^{(l)}$ with $\begin{gathered}\includegraphics[height=1cm]{GHZ_circle} \end{gathered}$ (primate circuit  digraphs).
	
To see	this, we first check that there is a one-to-one correspondence between the DCCs in a path circuit digraph and those in a sculpting digraph that replaces some of loops with primate circuits. Let us first choose $A_1$ for replacing its loop with primate circuit  digraphs of length $K$ as follows: 
		\begin{align}
		\includegraphics[width=.5\textwidth]{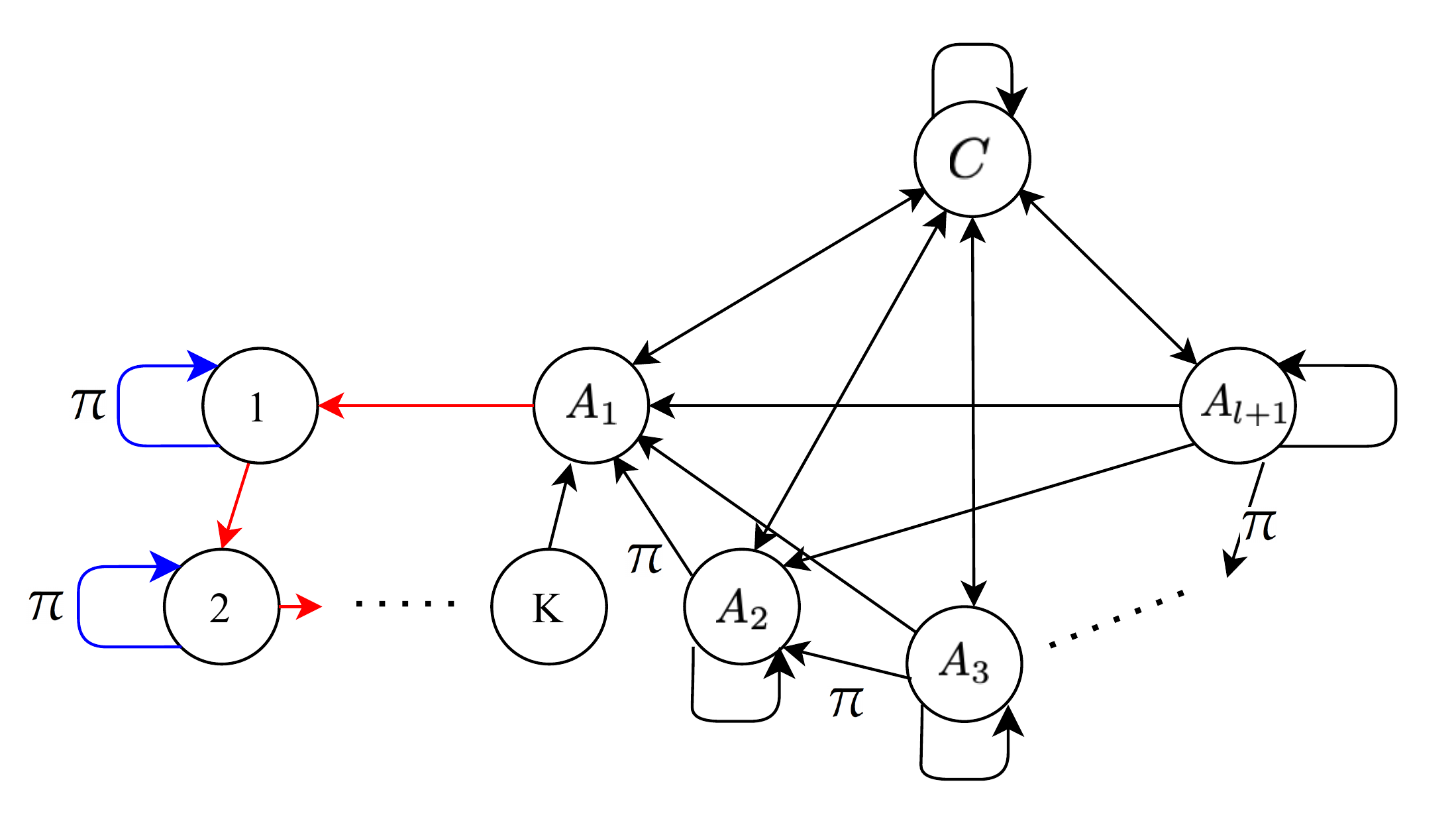} \nn 
		\end{align}
		Then any DCC with the replaced loop instead includes
		$\begin{gathered}\includegraphics[height=1cm]{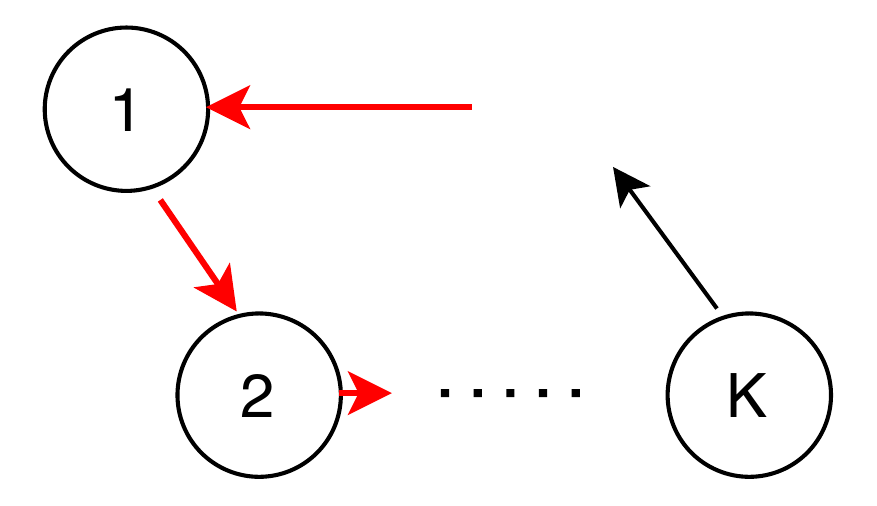} \end{gathered}$, and any DCC without the loop instead includes 
		$\begin{gathered}\includegraphics[height=1cm]{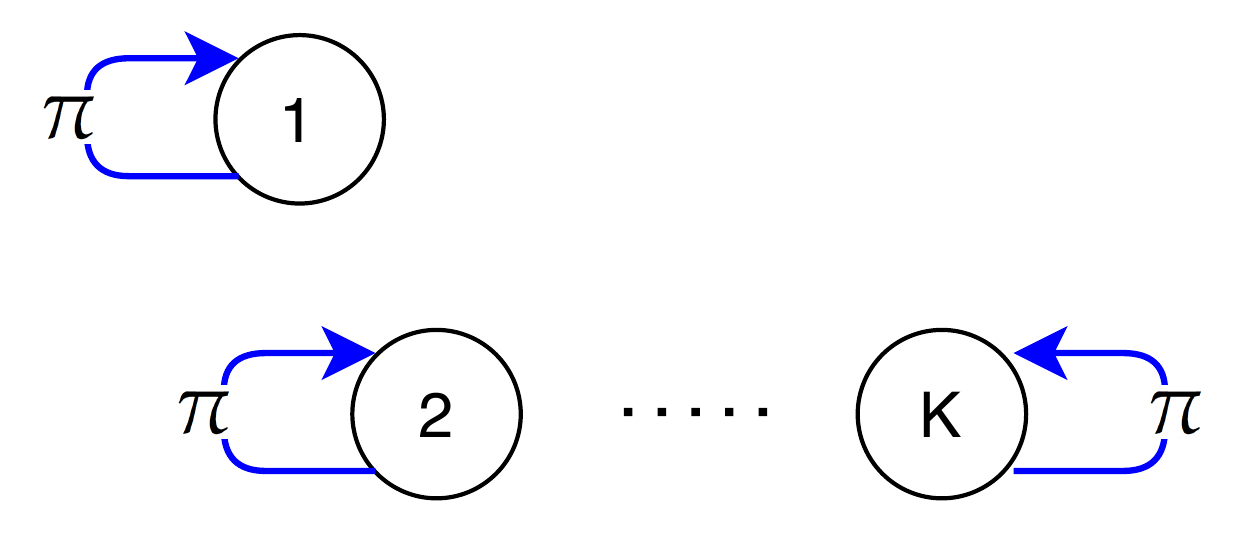} \end{gathered}$. The total number of DCCs is preserved under the replacement. The same is true with any replacements of loops on all $A_j$'s. 
This property implies that when $A_j$ ($j\in \{1,2,\cdots, {l+1}\}$) has  a loop in a directed PM of $P^{(l)}$, the state of $j_{k}$ ($k\in\{1,2,\cdots K_j\}$) is 1, and otherwise 0. 

Second, the number of DCCs for a given $P^{(l)}$ is $2^{l+1}$. This can be verified by using the fact that the number of directed PMs of a directed graph is given by the permanent of its unweighted adjacency matrix. Defining such an adjacency matrix as $P^{(l)}_{uw}$, we can see that Perm$[P^{(1)}_{uw}]$ (the matrix permanent of $P^{(1)}_{uw}$) is 4 and hence Perm$[P^{(l)}_{uw}]$= 2$\times$Perm$[P^{(l-1)}_{uw}] = 2^{l+1}$. This implies that the number of terms of the target state generated from the sculpting digraph is equal to $2^{l+1}$.  

Third, one can check that the submatrix of $P^{(l)}$ associated to $a_{A_jA_j}$ ($j\in\{1,2,\cdots, l+1\}$) has the same nonzero elements with $P^{(l-1)}$, which means that any subsystem connected to $A_j$ has the same number of 0 and 1 in all $2^{l+1}$ terms.  

Fourth, the distribution of edge weights $\pi$ in $P^{(l)}$ implies that a minus sign comes when the two neighboring system are in the state $1$, the essential property of the CZ gate. 

    
Combining the above four properties, we can see that the final state generated by the sculpting digraph that is a $P^{(l)}$ with all loops of $A_j$ ($j\in\{1,2,\cdots, l+1\}$) replaced with  primate circuits.

\section{Success probability calculation}\label{app:suc_prob}

We can calculate the success probability of our scheme by calculating the probability of obtaining the desired heralding outcome from the initial photonic state. This probability is obtained directly from the normalization condition: starting from a normalized input state, we identify the probability amplitude associated with the normalized target state for the desired heralding outcome. The success probability is then given by the squared modulus of this probability amplitude.

To perform the calculation, we divide the entire circuit in Fig.~\ref{fig:loc} into $l+1$ primate circuits and $P^{(l)}$.  

\begin{figure}
    \centering
    \includegraphics[width=0.8\linewidth]{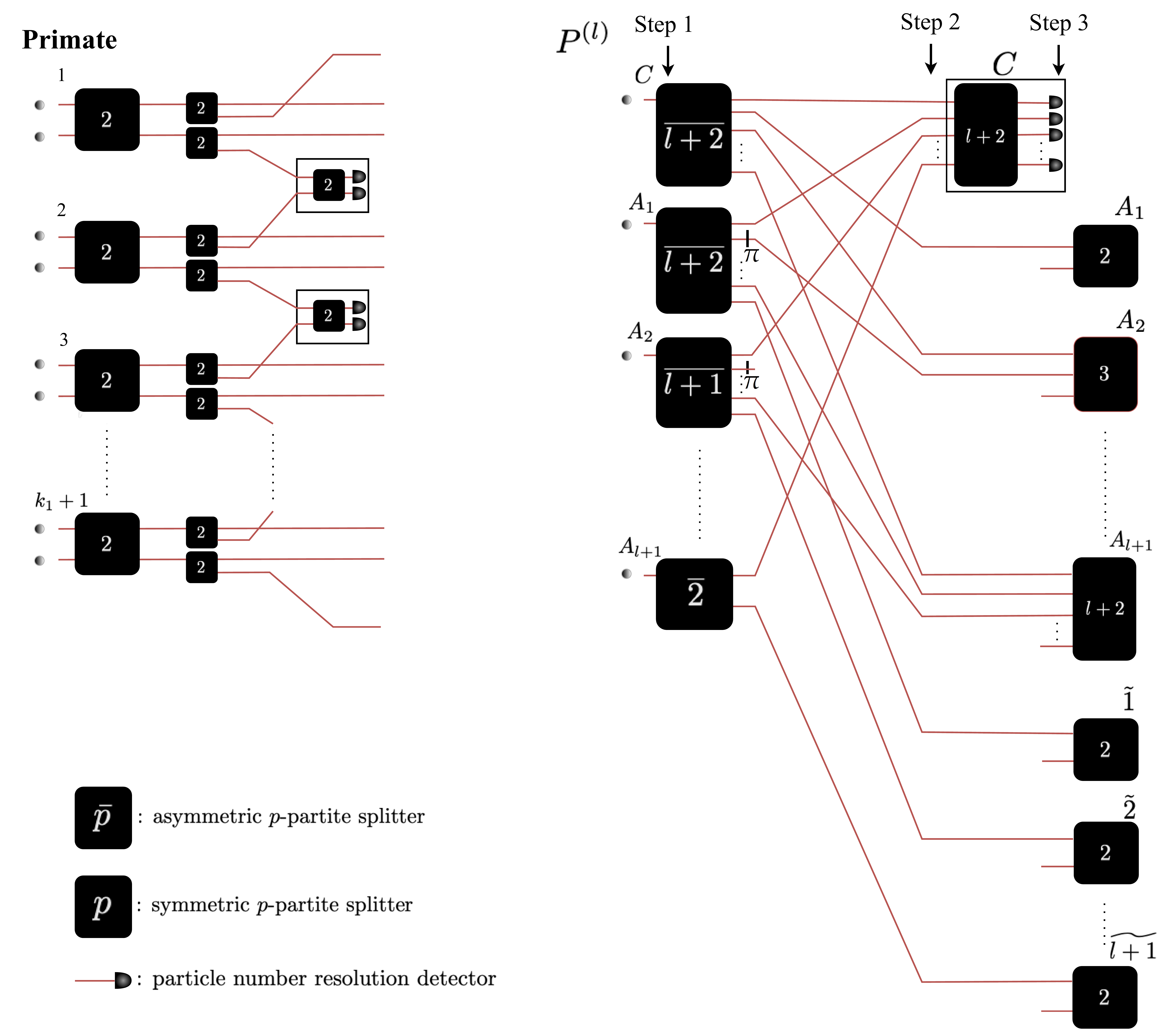}
    \caption{Decomposition of our CGS-generating circuit in Fig.~\ref{fig:loc} into $l+1$ primate circuits (left) and $P^{(l)}$ (right). }
    \label{fig:primate circuit _p_l}
\end{figure}

 First, the primate circuits defined here (see the left circuit of Fig.~\ref{fig:primate circuit _p_l}) are fundamentally identical to those in Ref.~\cite{bart2021creation}, except for the 2 splitters on rail 0 of the first mode and rail 1 of the last mode~(see Fig. 11 of Ref.~\cite{bart2021creation}). Therefore, from Eq. (4) of Ref.~\cite{bart2021creation}, the  state of primate circuit  1 after the postselections is directly written as
\begin{align}\label{eq:primate circuit _state}
\Big(\frac{1}{2}\Big)^{k_1+1} \Big[\big( \prod_{j=1}^{k_j}\ha^\dagger_{j0} \big)\ha^\dagger_{10'} +  \big( \prod_{j=1}^{k_j}\ha^\dagger_{j1} \big)\ha^\dagger_{k_j 1'} \Big] +|\chi\> 
\end{align} where $|\chi\>$ is the part that do not contribute to the final state after merged with $P^{(l)}$.  The other primate circuit  states are all written in the same form.

Second, $P^{(l)}$ can be divided into three steps of circuit operations (see the right circuit of Fig.~\ref{fig:primate circuit _p_l}). 
\begin{itemize}
\item Step 1: One photon is injected to the 0th rail of each mode $C, A_1,\cdots, A_{l+1}$, by which the normalized initial state is given by
\begin{align}
   |P^{(l)}_{\text{init}}\>  =   \ha^\dagger_{C0}\prod_{q=1}^{l+1}\had_{A_q 0}|vac\>.
\end{align}
\item Step 2: Each operators denoted as $\bar{p}$ ($2\le p \le l+2$) split a photon into $p$-rails, which is possible by combinations of unitary operators and vacuum~\cite{reck1994experimental}. After the beamsplitting and rewiring, the state evolves into    
\begin{align}
     (\sum_{p=1}^{l+1}\a_{p}^C\hb^\dagger_{p0}+\a_{0}^C \hb^\dagger_{C0} ) \Big[\prod_{j=1}^{l}\Big(\a_{j}^j\hb^\dagger_{\tilde{j}0} - \a_{j+1}^{j}\hb^\dagger_{j+1,1}+\sum_{p=j+2}^{l+1}\a_{p}^j\hb^\dagger_{p,p-2} +\a_0^j \hb^\dagger_{Cj} \Big)\Big]\Big( \a_{l+1}^{l+1}\hb^\dagger_{\widetilde{l+1},0}+\a_{0}^{l+1}\hb^\dagger_{C,l+1}  \Big).
    \end{align}
Note that the probability amplitudes from the left to right can be expressed as an $(l+2)\times (l+2)$ matrix as follows:
\begin{align}
A = 
\begin{pmatrix}
     \a^C_1 & \a^C_2 & \a^C_3 & \cdots & \a^C_{l+1} &a_0^C \\
 \a^1_1 & -\a^1_2 & \a^1_3 & \cdots & \a^1_{l+1}&a_0^1 \\
  0 & \a^2_2 & -\a^2_3 & \cdots & \a^2_{l+1} &a_0^2 \\ 
  0 & 0 & \a^3_3 & \cdots & \a^3_{l+1} &a_0^3 \\ 
 \vdots &\vdots & \vdots & \vdots &\vdots & \vdots \\
 0 & 0 & 0 & \cdots & \a^{l+1}_{l+1} &a_0^{l+1} \\   
\end{pmatrix} 
\end{align} We can set the elements of $A$ so that they give the maximal success probability.

\item Step 3: Each black box $p$ ($2\le p \le l+2$) is a $p\times p$ symmetric multiport splitter that executes the discrete Fourier transformation~\cite{lim2005generalized}. Then we postselect the events where only one detector in $C$ clicks and feed-forwards.
We can see that the probability amplitudes of all the non-zero terms are expressed as the nonzero terms of the matrix  permanent of $A$. From this observation, we impose two constraints on the elements of $A$:
\begin{enumerate}
\item All terms in the permanent of $A$ have the same absolute value. 
    \item  The 1st and 2nd rows of $A$ construct $(l+2)$-dimensional sphere and the $m$th row for $m \ge 3$ constructs a $(l+4-m)$-dimensional sphere. 
\end{enumerate}
The first constraint comes from the form of the CGSs. The second is from the normalization conditions.
From these constraints, $A$ is written as
 \begin{align}\label{app:asymmetric_amp}
\begin{pmatrix}
   \cos\theta_{l+1} &   \sin\theta_{l+1} \cos\theta_{l} & \cdots &\cdots & \prod_{j=4}^{l+1}\sin\theta_j\cos\theta_{3} & \prod_{j=3}^{l+1}\sin\theta_j\cos\theta_{2} & \prod_{j=2}^{l+1}\sin\theta_j\cos\theta_{1} & \prod_{j=1}^{l+1}\sin\theta_j & \\
 \cos\theta_{l+1} & -\sin\theta_{l+1} \cos\theta_{l} & \cdots &\cdots & \prod_{j=4}^{l+1}\sin\theta_j\cos\theta_{3} & \prod_{j=3}^{l+1}\sin\theta_j\cos\theta_{2} & \prod_{j=2}^{l+1}\sin\theta_j\cos\theta_{1} & \prod_{j=1}^{l+1}\sin\theta_j  \\
    0 & \cos\theta_l & \cdots  & \cdots &  \prod_{j=4}^{l}\sin\theta_j\cos\theta_{3} & \prod_{j=3}^{l}\sin\theta_j\cos\theta_{2} & \prod_{j=2}^{l}\sin\theta_j\cos\theta_{1} &  \prod_{j=1}^{l}\sin\theta_j  \\
 \vdots &\vdots & \vdots & \vdots &\vdots & \vdots &\vdots & \vdots  \\
  0 &0 & \cdots &0 & \cos\theta_3 & -\sin\theta_3\cos\theta_2 & \prod_{j=2}^{3}\sin\theta_j\cos\theta_{1} &\prod_{j=1}^{3}\sin\theta_j   \\ 
 0 &0 & \cdots &0 & 0 & \cos\theta_2 & -\sin\theta_2\cos\theta_1 &  \prod_{j=1}^{2}\sin\theta_j  \\ 
0 &0 & \cdots &0 & 0 & 0 &\cos\theta_1 &\sin\theta_1 \\ 
\end{pmatrix}     
 \end{align}

\end{itemize}

The amplitude of each non-zero term by the postselection is equal to 
\begin{align}\label{eq:p_l_suc}
    \prod_{j=1}^{l+1}\sin\theta_j\cos\theta_j = \Big(\frac{1}{2}\Big)^{l+1}\prod_{j=1}^{l+1}\sin2\theta_j.
\end{align}
We can directly see that the probability for successfully generating the $P^{(l)}$ state becomes maximal, i.e., $\big(\frac{1}{2}\big)^{2(l+1)}$,  when $\theta_j = \theta = \pi/4$. 

Finally, when all the primate circuits are connected to $P^{(l)}$ as in Fig.~\ref{fig:loc}, the entire success probability after the normalization is given by
\begin{align}
 \underbrace{\Big(\frac{1}{2}\Big)^{2\sum_{j=1}^{l+1}(k_j+1)}}_{\text{primate circuits}}\times \underbrace{\Big(\frac{1}{2}\Big)^{2(l+1)}}_{P^{(l)}}\times \underbrace{2^{l+1}}_{\text{target normalization}} = \Big(\frac{1}{2}\Big)^{2\sum_{j=1}^{l+1}k_j + 3l+3}.\end{align}
Here the primate circuit factor comes from Eq.~\eqref{eq:primate circuit _state}, the $P^{(l)}$ factor is determined from Eq.~\eqref{eq:p_l_suc}. The target normalization factor comes from the fact that a CGS has  $2^{l+1}$ equal-amplitude orthogonal terms,  so the squared common amplitude is multiplied by $2^{l+1}$ when projecting onto the normalized target state. 


\end{document}